\documentclass[prb,twocolumn,superscriptaddress]{revtex4-1}
 \usepackage{amsmath,amssymb, mathrsfs}
 \usepackage{stmaryrd}
 \usepackage{latexsym}
 \usepackage{CJK}
 \usepackage{graphicx}
 \usepackage{indentfirst}
 \usepackage{listings}
 \usepackage{xcolor}
 \usepackage{booktabs}
 \usepackage{stmaryrd}
 \usepackage{multirow}
 \usepackage{verbatim}
 \usepackage{makecell}
 \usepackage{lineno,hyperref}
 \usepackage{longtable}
 \usepackage{hyperref}
  \usepackage{upgreek}

\hypersetup{colorlinks,breaklinks,
            linkcolor=blue,urlcolor=blue,anchorcolor=blue,citecolor=blue}

\newcommand{\ud}{\mathrm{d}}

\newcommand{\ug}{\mathrm{g}}
\newcommand{\uc}{\mathrm{c}}

\newcommand{\uA}{\mathrm{A}}
\newcommand{\uB}{\mathrm{B}}

\newcommand{\sgn}{\,\text{sgn}}

\begin{document}

\title{Disconnection-Mediated Migration of Interfaces in Microstructures:\\
I. continuum model}

\author{Jian Han}
\affiliation{Department of Materials Science and Engineering, City University of Hong Kong, Hong Kong SAR, China}
\author{David J. Srolovitz}
\affiliation{Department of Mechanical Engineering, The University of Hong Kong, Pokfulam Road, Hong Kong SAR, China}
\author{Marco Salvalaglio}
\affiliation{Hong Kong Institute for Advanced Study, City University of Hong Kong, Hong Kong SAR, China}
\affiliation{Institute of Scientific Computing, TU Dresden, 01062 Dresden, Germany}
\affiliation{Dresden Center for Computational Materials Science, TU Dresden, 01062 Dresden, Germany}


\begin{abstract}
A long-standing goal of materials science is to understand, predict and control the evolution of microstructures in crystalline materials. 
Most microstructure evolution is  controlled by interface motion; hence, the establishment of rigorous interface equations of motion is a universal goal of materials science. 
We present a new model for the motion of arbitrarily curved interfaces that respects the underlying crystallography of the two phases/domains meeting at the interface and is consistent with  microscopic mechanisms of interface motion; i.e.,  disconnection   migration (line defects in the interface with step and dislocation character).
We  derive the equation of motion for  interface migration under the influence of a wide range of driving forces. 
In Part II of this paper [Salvalaglio, Han and Srolovitz, 2021], 
we  implement the interface model and the equation of motion proposed in this paper in a diffuse interface simulation approach for complex morphology and microstructure evolution. 
\end{abstract}

\maketitle

\section{Introduction}

Most materials of engineering significance are polycrystalline and/or multi-phase.
The resulting microstructures can be characterised by spatial distributions of domains of different phases (e.g. multi-phase alloys) or different crystallographic orientations (e.g. polycrystals). 
In this manner, the microstructure may be viewed as a network of interfaces 
and  microstructure evolution as the motion of these interfaces. 
The key to such evolution is the governing interface equation of  motion. 
Here, we present such an interface equation of  motion in crystalline systems that respects the underlying crystal structure(s) of the materials, the bicrystallography of their interfaces, and is suitable for incorporation into complex, microstructure evolution simulations. 

The formulation of such a governing equation relies on crystallography-respecting interface models and the mechanisms by which they move.  
The simplest model is to treat an interface as a smoothly curved, homogeneous, isotropic surface that moves as dictated by the reduction of the total free energy of the system (gradient flow). 
If the only contribution to the free energy is the excess energy of the interfaces (surface tension), interface motion is classical curvature flow~\cite{mullins1956two,burke1952recrystallization}: 
{\color{black}$\mathbf{v} = M\gamma_0 \kappa \hat{\mathbf{n}}$, where $v$ is the local interface velocity, $M$ is its (constant) mobility, $\gamma_0$ is the (constant) interface energy, $\kappa$ is the local mean curvature, and $\hat{\mathbf{n}}$ is the local normal of the interface plane.} 
This model and governing equation can be easily implemented in large-scale microstructure evolution simulations (e.g. normal grain growth simulations~\cite{lazar2010more}). 
For an anisotropic surface (i.e., surface orientation-dependent properties), we only need to replace the interface energy $\gamma_0$ by the interface stiffness~\cite{herring1951physics,du2007properties}. 

Such a continuum interface formulation  is simple but ignores the role of crystallography on interface motion. 
A simple, crystallography-respecting interface approach is the terrace-ledge-kink (TLK) model which was developed to describe the free surfaces of crystals~\cite{kossel1927theorie,burton1951growth,mullins1963microscopic}. 
The TLK model was  extended to grain boundaries (GBs) half a century ago~\cite{gleiter1969mechanism}. 
In this model, one interface orientation is considered a reference (or terrace) such that, at the continuum scale, the interface morphology can be described by a single-valued height function $z(x,t)$ measured from the reference. 
The kinetic mechanism for interface motion in the TLK model is the propagation of steps (i.e., ledges and kinks) along the terrace. 
A continuum version of the step-flow picture was previously proposed~\cite{stone2005continuum}.

Models originally developed for crystal surfaces, such as the TLK model, cannot be directly applied to interfaces in crystalline materials. 
This is because  interfacial line defects are not pure steps/ledges; they are, in general, characterised by both a step height and a Burgers vector; these line defects are {\it disconnections}~\cite{bollmann1970general,ashby1972boundary,hirth1973grain}. 
Disconnections have been widely observed on GBs~\cite{rajabzadeh2013evidence,zhu2019situ} and heterophase interfaces~\cite{pond2003comparison,zheng2018determination,qi2020interdiffusion}. 
The Burgers vector character of disconnections plays a significant effect in the motion of interfaces; including the following.
(i) An applied stress can drive the glide of disconnections;  the step character necessarily couples this to interface migration  (i.e., shear-coupled migration~\cite{cahn2006coupling}). 
(ii) When an interface migrates, the propagation of  finite Burgers vector disconnections necessarily leads to  shear across the interface, such as in Martensitic transformations~\cite{bhattacharya2003microstructure}. 
(iii) In a microstructure where the interfaces are highly curved and/or interconnected,  shear along individual interface will be constrained such that internal stresses will accumulate during  microstructure evolution; this implies that microstructure evolution may stagnate~\cite{thomas2017reconciling}. 
(iv) Elastic interactions amongst  disconnections suggest that interfaces may undergo a  Kosterlitz-Thouless transition at a critical temperature~\cite{chen2020grainb} (signalled by an abrupt change in the temperature dependence of GB mobility and viscosity). 
(v) Elastic interactions amongst disconnections  lead to GB thermal fluctuation spectra that follow $k^{-1}$ rather than $k^{-2}$ (like free surfaces and solid-fluid interfaces), where $k$ is the wave vector~\cite{karma2012relationship}. 

\begin{figure*}[tb]
\includegraphics[height=0.6\linewidth]{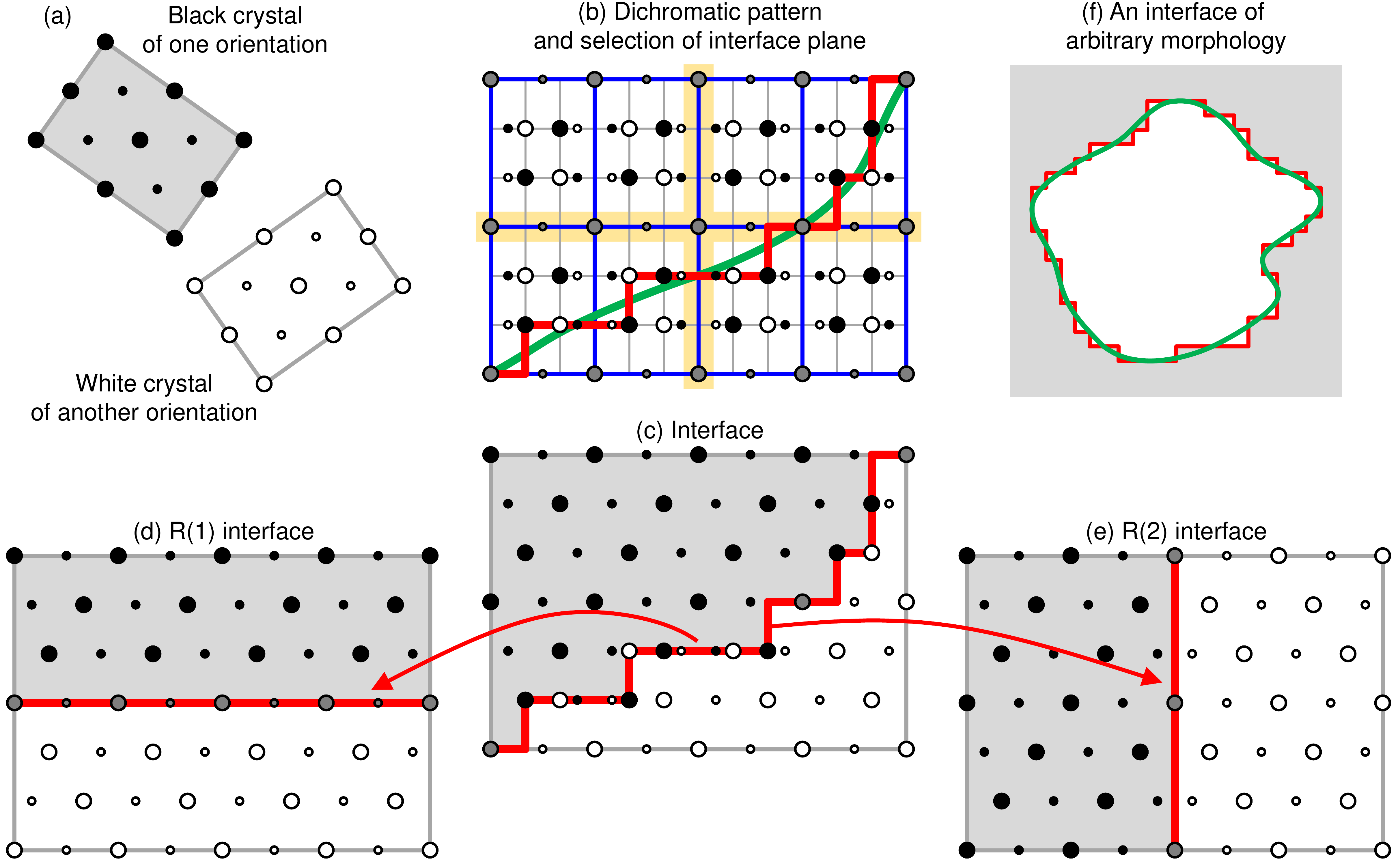}\hspace{-1.78em}%
\caption{\label{fig:DSC}
Schematic of the two-reference interface model for a $\Sigma 3$ $[100]$ $(11\bar{2})$ symmetric tilt grain boundary in a body-centred cubic crystals  
(a) Two crystals (black and white) of different orientations and/or lattice structures.
(b) Dichromatic pattern formed by extending two crystals throughout  space (blue and grey grids denote the CSL and the DSC lattices). 
The green curve is a continuum representation of the interface;  the red curve corresponds to its discretisation on the DSC lattice. 
The first and second closest-packed planes are highlighted. 
(c) The interfaces corresponding to (b). 
(d) and (e) show the two reference interfaces. 
From (d) to (c) to (e), the interface inclination angle varies from 0 to $90^\circ$. 
(f) A continuum interface for an arbitrary morphology domain (green curve) described in the two-reference description (red curve). 
}
\end{figure*}

We recently proposed a disconnection glide-based, continuum equation of motion for GBs~\cite{zhang2017equation}. 
Similar to the TLK model, it is limited by an intrinsic assumption; i.e., there is a special, flat reference interface and the interface morphology represents only small deviations from this reference interface. 
However, in a microstructure, interfaces are commonly highly curved such that the assumption of a small deviation from a reference interface is inappropriate and prevents its application in general microstructure evolution simulations {\color{black}or very wide boundaries}. 
In this paper, we propose a continuum two-reference interface equation of motion for arbitrarily curved interface of arbitrary bicrystallography. 
This paper is organised as follows. 
In Sect.~\ref{theory}, we develop a theory for interface kinetics and then, in Sect.~\ref{numerical} we apply this to sharp interface simulations for several simple cases. 
In Sect.~\ref{discussion}, we discuss the physical meaning of several parameters in the model and provide its extensive generalisation.
In Part II of this paper~\cite{Salvalaglio2021}, 
we develop a diffuse interface approach for the simulation of the evolution of an interface or multiple interfaces in complex settings.

\section{Theory of interface dynamics}\label{theory}

\subsection{Dual-reference interface model}\label{theorytworef}

We describe a general interface between a pair of crystals in two steps\cite{SuttonBalluffi,han2018grain}:
\begin{itemize}
\item[(i)]
Represent each crystal by its translation vectors and rotate one lattice with respect to the other to the target misorientation.
Interpenetrate the two crystal lattices to form a dichromatic pattern and the associated coincidence-site lattice (CSL). 
For example, Fig.~\ref{fig:DSC}b shows the dichromatic pattern formed by the black and white crystals of the same type, Fig.~\ref{fig:DSC}a. 
The grey grid in Fig.~\ref{fig:DSC}b denotes the displacement-shift-complete (DSC) lattice, while the blue grid denotes the CSL. 

\item[(ii)]
Select an interface plane. 
Remove one crystal (black lattice points) on one side of the interface and the other crystal (white lattice points) on the other side (see Figs.~\ref{fig:DSC}b and c). 
\end{itemize}

While there is ambiguity in choosing the interface plane, close-packed CSL planes are special (see the first and second closest-packing planes  highlighted in Fig.~\ref{fig:DSC}b); special in the sense that they are highly coherent and often associated with cusps in  interface energy versus inclination plots and are often micro- (or macro-) facets.
We choose these  planes as the two reference interfaces, denoted by ``R(1)'' and ``R(2)'' in Figs.~\ref{fig:DSC}d and e. 
These special interfaces are ``reference'' interfaces in the following senses.  
An interface with a plane slightly inclined with respect to R(1) (Fig.~\ref{fig:DSC}d) can be described as an R(1) interface with a superimposed distribution of disconnections (line defects in the interface plane with both a step height and a Burgers vector each corresponding to DSC lattice vectors -- see Fig.~\ref{fig:DSC}b). 
Similarly, an interface that is slightly inclined with respect to an R(2) interface (Fig.~\ref{fig:DSC}e) can be described as an R(2) interface with superimposed disconnections of another type. 
This description of inclined interfaces is well-established~\cite{Sutton83b}.
{\color{black}While Fig.~\ref{fig:DSC} illustrates a particularly simple case (corresponding to a $\Sigma$3 [100] (11$\bar{2}$) symmetric tilt grain boundary in a body-centred
cubic crystal), the Supplement Material shows two additional examples of reference interface selection in more complex  DSC lattices.}

Generalisation of this concept requires a description of interfaces for arbitrary inclinations with respect to any plane.
Such interfaces are  an intrinsic features of all non-faceted microstructures. 
To address this issue, we introduce a two-reference model as follows. 
First, we assume that a disconnection on the R(1) interface can be viewed as a segment of the R(2) interface and vice versa. 
This assumption is consistent with the myriad observation from atomistic simulations and experiments~\cite{dodaran2019energetic,medlin2017defect}.
In this representation, an arbitrarily curved interface can be decomposed into segments of R(1) and R(2) interfaces, as illustrated in Fig.~\ref{fig:DSC}c. 
Accordingly, an interface can be described by two sets of disconnections; one on each of the  reference interfaces. 
The shape of the interface is described by the distribution of disconnection-steps while the stress field (relative shear displacements) near the interface is caused by the distribution of disconnection-Burgers vectors. 

Since the reference interfaces should be close-packed CSL planes, the two reference interfaces need not be perpendicular to each other; this depends on crystallography. 
However, for concreteness and simplicity of representation, we present the two-reference model for a situation in which the two reference interfaces are orthogonal.
This applies, for example, to the case of $[110]$ tilt grain boundaries in cubic crystals (Fig.~\ref{fig:DSC}). 
The extension to the non-orthogonal case is discussed in Sect.~\ref{discussion}.

\subsection{Pure-step disconnection}\label{stepmodel}

\begin{figure}[t!]
\includegraphics[height=1.2\linewidth]{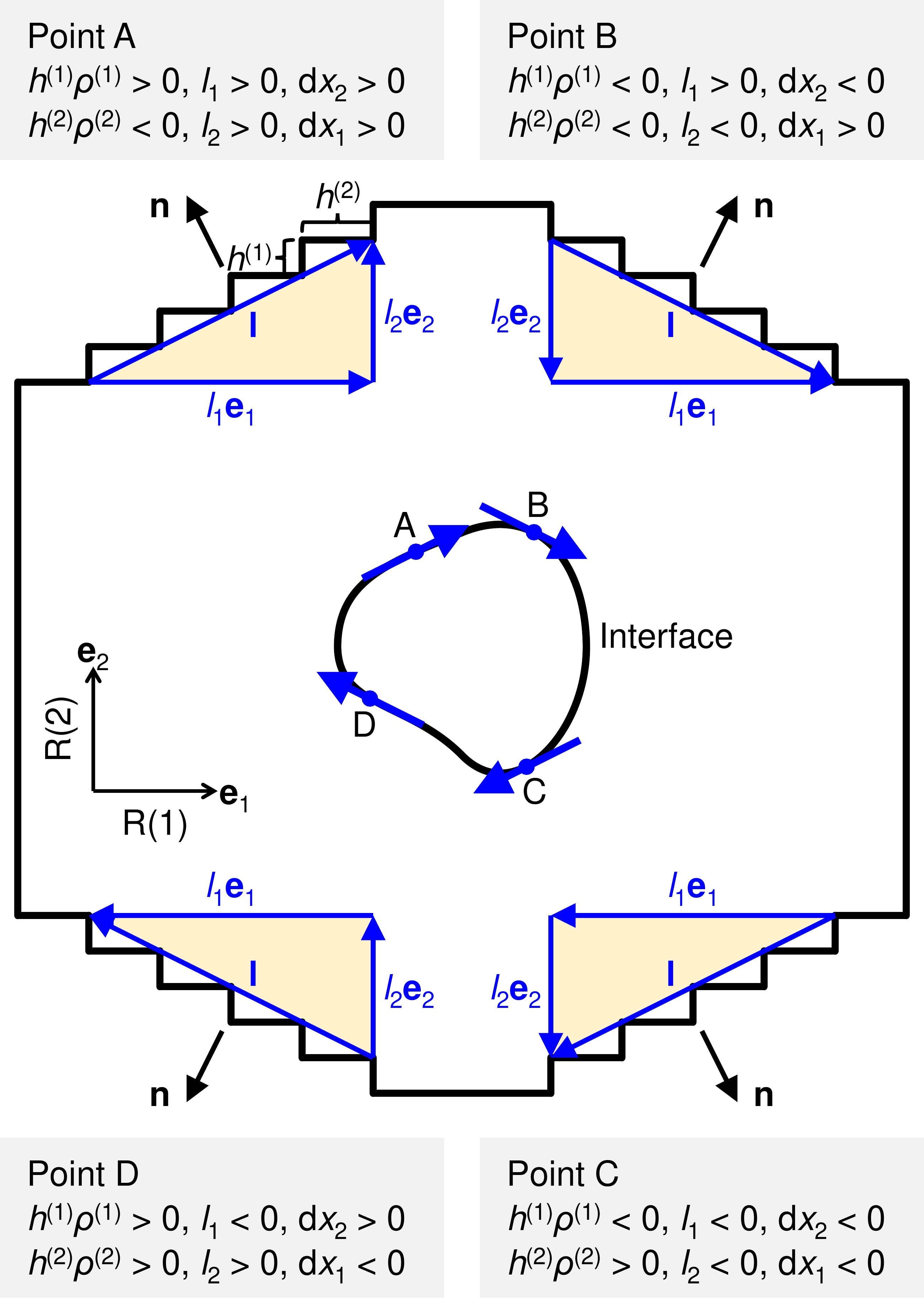}\hspace{-1.78em}%
\caption{\label{fig:tangentgeometry}
A closed interface curve (centre) and its piecewise representation in the reference lattice  (outer). 
The tangent vector points clockwise. 
The steps on the four segments of arc length $|\mathbf{l}|$ at Points A, B, C and D are shown explicitly. 
The geometry is consistent with Eqs.~\eqref{h1rho1t2} and \eqref{h2rho2t1}. 
}
\end{figure}

Although each segment of an interface (e.g., red lines in Fig.~\ref{fig:DSC}b) can be characterised, in general, by step heights and Burgers vectors on the two reference interfaces, we first present a simple case in which the Burgers vectors are zero; i.e., a pure-step model (see Fig.~\ref{fig:tangentgeometry}). 

\subsubsection{Geometric description}
An arbitrary curve in the $\mathbf{e}_1$-$\mathbf{e}_2$ space (Fig.~\ref{fig:tangentgeometry}) can be expressed parametrically as
\begin{equation}
\mathbf{x}(s) = \left(
\begin{array}{c}
x_1(s) \\ x_2(s)
\end{array}
\right), 
\end{equation}
where $s$ is the parameter. 
The local tangent vector $\mathbf{l}$   is 
\begin{subequations}\label{tangent}
\begin{gather}
\mathbf{l}
= \frac{\ud \mathbf{x}}{\ud s}
= \left(\begin{array}{c}
\ud x_1/\ud s \\ \ud x_2/\ud s
\end{array}\right)
=
\left(
\begin{array}{c}
l_1 \\ l_2
\end{array}\right),
\text{ or}
\end{gather}
\begin{gather}
\hat{\mathbf{l}}
= \frac{\mathbf{l}}{|\mathbf{l}|}
= \frac{1}{|\mathbf{l}|}\frac{\ud\mathbf{x}}{\ud s}
= \frac{\ud\mathbf{x}}{\ud L},
\end{gather}
\end{subequations}
where $\hat{\mathbf{l}}$ is the normalised tangent  and the change in arc length $\ud L$ corresponding to a unit change in $s$ is $|\mathbf{l}|\,\ud s$. 

Define $\rho^{(k)}(s) \,\ud L$ ($k=1,2$) as the number of R($k$) disconnections in the arc $\ud L$; 
{\color{black}in other words, $\rho^{(k)}$ represents the number of R($k$) disconnections per unit arc length. }
$\rho^{(1)} > 0$ if, when we go along the $\mathbf{e}_1$ direction, R(1) step leads to  increment $\ud x_2 > 0$ along the $\hat{\mathbf{l}}$ direction; and vice versa. 
Similarly, $\rho^{(2)} > 0$ if, when we go along the $\mathbf{e}_2$ direction, R(2) step leads to  increment $\ud x_1 < 0$ along the $\hat{\mathbf{l}}$ direction. 
R(1) and R(2) disconnections contribute to an increment of $x_2$ ($x_1$) by 
\begin{subequations}
\begin{gather}\label{h1rho1t2}
h^{(1)} \rho^{(1)} \ud L 
= \ud x_2
\Rightarrow
h^{(1)} \rho^{(1)} 
= \hat{l}_2, 
\end{gather}
\begin{gather}\label{h2rho2t1}
- h^{(2)} \rho^{(2)} \ud L
= \ud x_1
\Rightarrow
- h^{(2)} \rho^{(2)} 
= \hat{l}_1,  
\end{gather}
\end{subequations}
respectively, where $h^{(k)}$ is the step height of the R($k$) disconnection.  
Equations~\eqref{h1rho1t2} and \eqref{h2rho2t1} give  
\begin{equation}\label{lrho}
\hat{\mathbf{l}}(s)
= \left(\begin{array}{c}
-h^{(2)} \rho^{(2)} \\
h^{(1)} \rho^{(1)}
\end{array}\right),
\end{equation}
which expresses the unit tangent vector in terms of $\rho^{(k)}(s)$ and can be integrated to find the  interface curve: 
\begin{equation}\label{trajectory}
\mathbf{x}(s)
= 
\left(
\def\arraystretch{2.2}
\begin{array}{c}
x_1(0) - \displaystyle{h^{(2)} \int_0^s \rho^{(2)}(s') |\mathbf{l}(s')|\,\ud s'}  \\ 
x_2(0) + \displaystyle{h^{(1)} \int_0^s \rho^{(1)}(s') |\mathbf{l}(s')|\,\ud s'} 
\end{array}\right). 
\end{equation}

We can check this result for a special case. 
For a flat, inclined interface plane with constant tangent unit vector $\hat{\mathbf{l}}$, $\rho^{(1)} = \hat{l}_2/h^{(1)}$ and $\rho^{(2)} = -\hat{l}_1/h^{(2)}$ are constant everywhere. 
For a flat R(1) interface (zero inclination) with $\hat{l}_1 = 1$ (i.e., the interface between Points A and B in Fig.~\ref{fig:tangentgeometry}), $\hat{l}_2 = 0$, $\rho^{(1)} = 0$ and $\rho^{(2)} = - 1/h^{(2)}$. 
$\rho^{(1)}$ and $\rho^{(2)}$ are not independent. 
When R(1) steps glide along the $\mathbf{e}_1$-axis, R(2) steps  simultaneously glide along  $\mathbf{e}_2$; the motions of the steps are simply two aspects of a single interface migration event. 
If so, why do we introduce a two-reference model? 
As we demonstrate below, the two-reference model provides a natural means of introducing anisotropy in interface properties (e.g., interface energy and mobility) along with a well-posed description of any interface orientation. 
The two-reference model allows us to associate a Burgers vector with each step that may be parallel to either the $\mathbf{e}_1$- or $\mathbf{e}_2$-axis.

\subsubsection{Curvature-flow kinetics}
\begin{figure}[b!]
\includegraphics[height=0.35\linewidth]{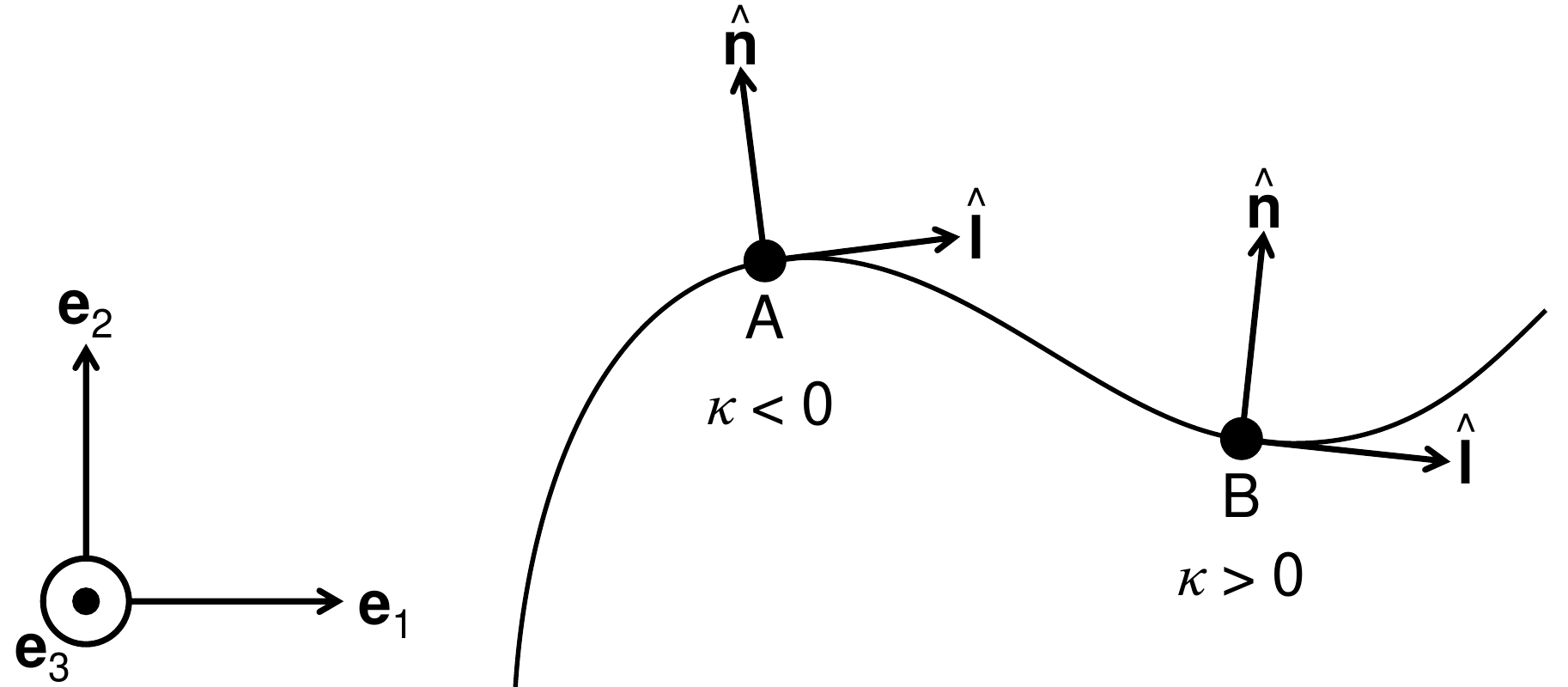}\hspace{-1.78em}%
\caption{\label{fig:convention}A 1D curve  in the $\mathbf{e}_1$-$\mathbf{e}_2$ plane. 
The normal  $\hat{\mathbf{n}}$ is defined such that $\hat{\mathbf{l}}\times\hat{\mathbf{n}} = \mathbf{e}_3$ where $\hat{\mathbf{l}}$ is the tangent. 
The curvature at Point A is $\kappa < 0$ while that at Point B is $\kappa > 0$. 
}
\end{figure}

In the pure-step case, the interface energy/surface tension is associated with both reference states and steps (localised energy). 
The total energy is a functional of the interface curve:
\begin{equation}\label{E}
E[\mathbf{x}(s)] = \int \gamma(s) |\mathbf{l}(s)| \,\ud s,  
\end{equation}
where the interface energy is a function of the local tangent direction: $\gamma(s) = \gamma(\hat{\mathbf{l}}(s))$ (in 2D it is a function of the inclination angle $\phi(s)$, defined relative to the R(1) interface). 
The driving force is the variation of the energy with respect to the interface displacement: 
\begin{equation}\label{capillarity}
\mathbf{f}(s)
= -\frac{\delta E[\mathbf{x}(s)]}{\delta \mathbf{x}(s)}
= \Gamma \kappa \left(
\begin{array}{c}
-\hat{l}_2 \\ \hat{l}_1
\end{array}\right)
= \Gamma \kappa \hat{\mathbf{n}}, 
\end{equation}
where $\Gamma \equiv \gamma + \gamma_{,\phi\phi}$ is the interface stiffness~\cite{herring1951physics,du2007properties}, $\kappa$ is the local curvature, and $\hat{\mathbf{n}} = (-\hat{l}_2, \hat{l}_1)^T$ is the local normal (see Fig.~\ref{fig:convention}). 
At Point A in Fig.~\ref{fig:convention}, $\kappa < 0$, $\mathbf{f}$ is in the  $-\hat{\mathbf{n}}$ direction; at Point B,  $\kappa > 0$ and $\mathbf{f}$ is in the $\hat{\mathbf{n}}$ direction. 
Equation~\eqref{capillarity} shows that the driving force is simply the classical capillary force (i.e., the weighted mean curvature\cite{taylor1992overview}), always directed along the normal. 
{\color{black}The Supplemental Material also shows how to represent the weighted mean curvature in terms of the interface divergence of capillarity vector \cite{hoffman1972vector}. }

In 2D, $\gamma = \gamma(\hat{\mathbf{l}})=\gamma(\phi)$. 
If the interface energy is inclination-independent/isotropic, $\Gamma = \gamma=\gamma_0$ and $\mathbf{f}(s) = \gamma_0 \kappa \hat{\mathbf{n}}$. 
The overdamped interface velocity is 
\begin{equation}\label{simplest}
\dot{\mathbf{x}}(s)
= M \mathbf{f}(s)
= M \gamma_0 \kappa \hat{\mathbf{n}}, 
\end{equation}
where $M$ is the interface mobility; 
{\color{black}here, we assume that $M$ is not a function of crystallography.} 
As a results, Eq.~\eqref{simplest} reduces to classical (isotropic) capillarity-driven interface migration~\cite{mullins1956two,burke1952recrystallization}. 

\begin{figure}[bt!]
\includegraphics[height=0.55\linewidth]{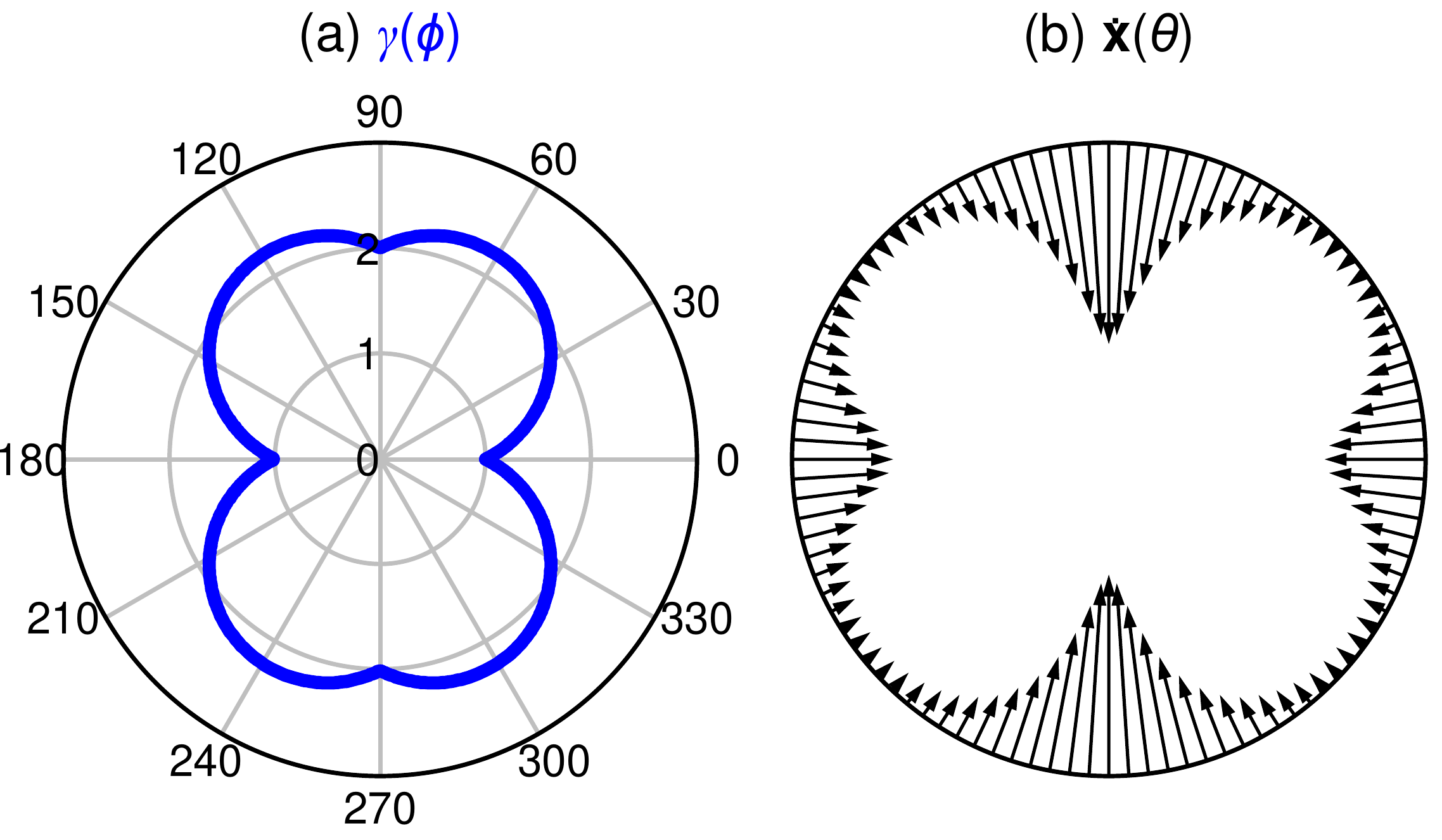}\hspace{-1.78em}%
\caption{\label{fig:anisotropy}(a) Polar plot of interface energy vs. inclination angle $\gamma(\phi)$ (scaled by $\gamma^{(2)}$) with $\gamma^{(1)}/\gamma^{(2)} = 2$. 
This plot corresponds to the $\gamma(\hat{\mathbf{n}})\hat{\mathbf{n}}$ plot in Ref.~\onlinecite{balluffi2005kinetics}. 
(b) The velocity field  around a circular interface (Eq.~\eqref{capillarity_anisotropy_regularisation} with $\epsilon = 0.3$ and $\gamma^{(1)}/\gamma^{(2)} = 2$). 
}
\end{figure}

The pure-step model gives rise to an anisotropic interface energy
\begin{equation}\label{gammadensity}
\gamma(s)= \gamma^{(2)} h^{(2)} |\rho^{(2)}(s)|+ \gamma^{(1)} h^{(1)} |\rho^{(1)}(s)|
\end{equation}
which, exploiting Eq.~\eqref{lrho} and the definition of $\phi$, implies
\begin{equation}\label{gammaaniso}
\gamma(\phi)= \gamma^{(2)} |\cos\phi| + \gamma^{(1)} |\sin\phi|,
\end{equation}
with $\gamma^{(k)}$ the ``step energy'' on the R($k$) interface (rather than interface energy). 
This interface energy encodes symmetries of the reference system~\cite{taylor1992overview} (see Sect.~\ref{sec:non-orth}). 
In this case, the interface energy has cusps at $\phi = n\pi/2$ ($n = 0, 1, 2, 3$)  corresponding to  R(1) and R(2) facets (e.g., see Fig.~\ref{fig:anisotropy}a). 
{\color{black}
Note that the geometry dependence of $\gamma$ is determined by the choice of reference interfaces and thus rooted in the crystallography, rather than arbitrarily chosen. 
As demonstrated in atomistic simulations~\cite{bulatov2014grain}, the form of $\gamma(\phi)$ (grain boundary energy) is indeed crystallography-determined. 
}
The $\gamma(\phi)$ in Eq.~\eqref{gammaaniso} implies the interface velocity 
\begin{equation}\label{capillarity_anisotropy}
\dot{\mathbf{x}}(s)= M \left[2\gamma^{(2)} \hat{l}_2^2 \delta(\hat{l}_1) 
+ 2\gamma^{(1)} \hat{l}_1^2 \delta(\hat{l}_2)\right] \kappa\hat{\mathbf{n}}, 
\end{equation}
where $\delta(x)$ is the Dirac delta function, which we regularise  as
\begin{equation}\label{Poissonkernel}
\delta(x) = \lim_{\epsilon\to 0} \frac{1}{\pi} \frac{\epsilon}{\epsilon^2 + x^2}
\end{equation}
such that Eq.~\eqref{capillarity_anisotropy} becomes
\begin{equation}\label{capillarity_anisotropy_regularisation}
\dot{\mathbf{x}}(s)
= M \frac{2\epsilon}{\pi} 
\left(
\gamma^{(2)} \dfrac{\hat{l}_2^2}{\epsilon^2 +\hat{l}_1^2} 
+ \gamma^{(1)} \dfrac{\hat{l}_1^2}{\epsilon^2 + \hat{l}_2^2}
\right)
\kappa \hat{\mathbf{n}},
\end{equation}
where $\epsilon$ is a small, dimensionless parameter (see Sect.~\ref{discussion_singularities}).
Consider the special case of a circular interface (radius $R$).
Here, $s = -\theta$ ($\hat{\mathbf{l}}$  and $\theta$ increase in the clockwise and counter-clockwise directions, respectively),  $x_1 = R\cos\theta$ and $x_2 = R\sin\theta$, $|\mathbf{l}| = \sqrt{x_1'^2 + x_2'^2} = R$, $\hat{l}_1 = x_1'/|\mathbf{l}| = \sin\theta$, $\hat{l}_2 = x_2'/|\mathbf{l}| = -\cos\theta$,  
$\kappa = -1/R$, and
$\hat{\mathbf{n}} = (-x_2', x_1')^T/\sqrt{x_1'^2 + x_2'^2} = (\cos\theta, \sin\theta)^T$  (see Fig.~\ref{fig:convention}).  
Then, Eq.~\eqref{capillarity_anisotropy_regularisation} reduces to the $\dot{\mathbf{x}}(\theta)$ shown in Fig.~\ref{fig:anisotropy} for $\epsilon = 0.3$ and $\gamma^{(1)}/\gamma^{(2)} = 2$. 

Figure~\ref{fig:anisotropy}b shows that the interface segment at $\theta = \pm\pi/2$ has the largest velocity; we can understand this as follows. 
This interface consists of R(1) steps of energy $\gamma^{(1)}$. 
Since $\gamma^{(1)}>\gamma^{(2)}$, the steps on the R(1) interface tend to be eliminated quickly.
This implies that the step density on the R(1) interface decreases and this interface segment becomes increasingly flat.
This result is counterintuitive since this interface segment ($\theta = \pm\pi/2$) has lower energy than nearby segments (e.g., $\theta = \pi/4$); if the interface velocity is simply proportional to interface energy, the higher-energy segments should migrate faster. 
Recall, however, that the anisotropic driving force is proportional to $\Gamma$ rather than $\gamma$; here $\gamma_{,\phi\phi}$ dominates the $\gamma$-contribution to $\Gamma$. 
At $\theta = \pi/2$ ($\phi = 0$), small deviation of $\phi$ from $0$ quickly increases the energy, implying  the interface flattens quickly. 

\subsubsection{Step-flow kinetics}
\begin{figure}[b!]
\includegraphics[height=0.35\linewidth]{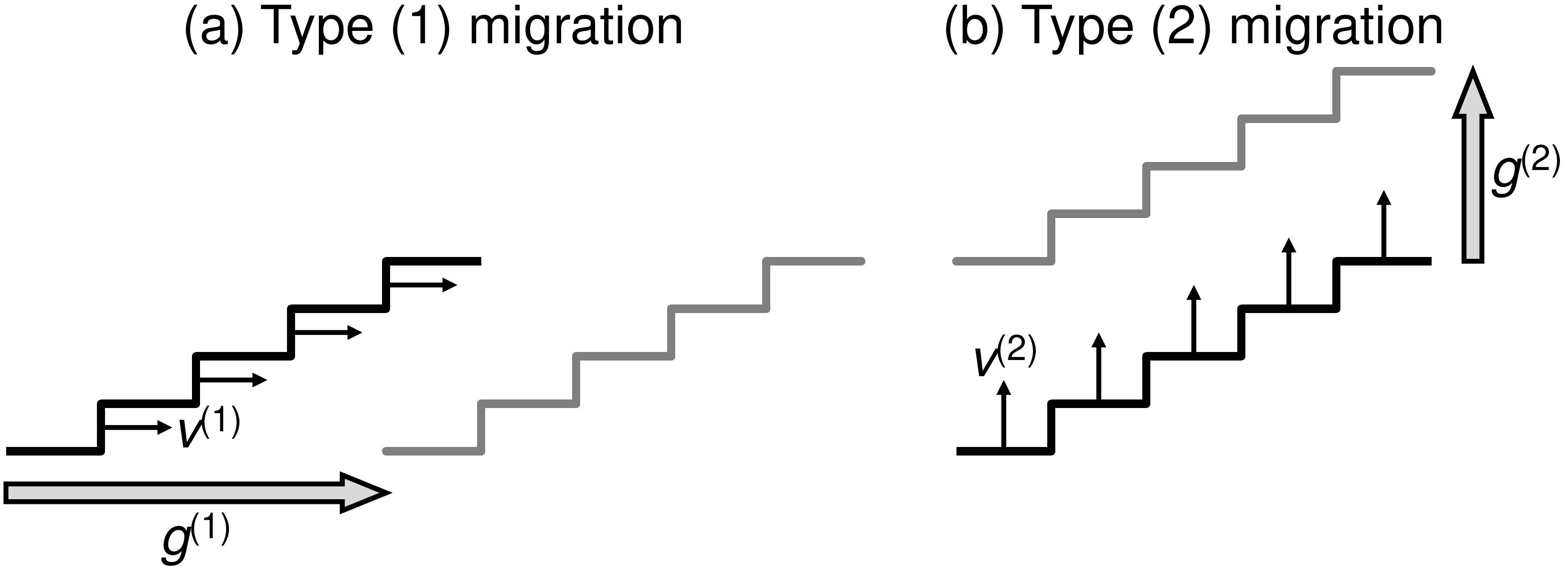}\hspace{-1.78em}%
\caption{\label{fig:migrationmode}
(a) Type~(1) migration is driven by  horizontal glide of R(1) steps. 
(b) Type~(2) migration is driven by  vertical glide of R(2) steps. 
The R($k$) step glide distance is $g^{(k)}$. 
}
\end{figure}

Here, we develop an equation of motion for the interface based on step motion. 
We distinguish two types of interface migration, as depicted in Fig.~\ref{fig:migrationmode}. 
Type~(1) migration (Fig.~\ref{fig:migrationmode}a) results from the driven glide of R(1) steps along the $\mathbf{e}_1$-axis; note that, although the R(2) steps are not driven, their motion is coupled. 
Similarly, Type~(2) migration (Fig.~\ref{fig:migrationmode}b) results from the driven glide of R(2) steps along the $\mathbf{e}_2$-axis, along with the coupled glide of R(1) steps. 

The glide distance and velocity of R($k$) steps along the $\mathbf{e}_k$-axis as $\mathbf{g}^{(k)}(s) = g^{(k)}(s) \mathbf{e}_k$  and $\mathbf{v}^{(k)}(s) = \dot{\mathbf{g}}^{(k)}(s) = v^{(k)}(s) \mathbf{e}_k$. 
We represent $\delta\mathbf{x}(s)$ (Eq.~\eqref{capillarity}) by $\delta g^{(k)}(s)$: 
\begin{equation}\label{dxtodg}
\delta \mathbf{x} 
= \left(\begin{array}{c}
\delta g^{(1)} \\ \delta g^{(2)}
\end{array}\right);
\end{equation}
i.e., interface migration can be decomposed into the two types of migration depicted in Fig.~\ref{fig:migrationmode}. 
The driving forces (per unit interface length) for the two types of interface migration are
\begin{equation}\label{capillarity_on_step}
f^{(1)}
= -\frac{\delta E}{\delta g^{(1)}}
= - \Gamma \kappa \hat{l}_2, \quad
f^{(2)}
= -\frac{\delta E}{\delta g^{(2)}}
= \Gamma \kappa \hat{l}_1. 
\end{equation}
Assuming overdamped step dynamics, the step velocities are
\begin{equation}\label{vstep}
v^{(1)} = - M^{(1)} \Gamma \kappa \hat{l}_2, \quad
v^{(2)} = M^{(2)} \Gamma \kappa \hat{l}_1, 
\end{equation}
where $M^{(k)}$ is the mobility associated with Type~($k$) step migration. 

\begin{figure}[bt!]
\includegraphics[height=1.1\linewidth]{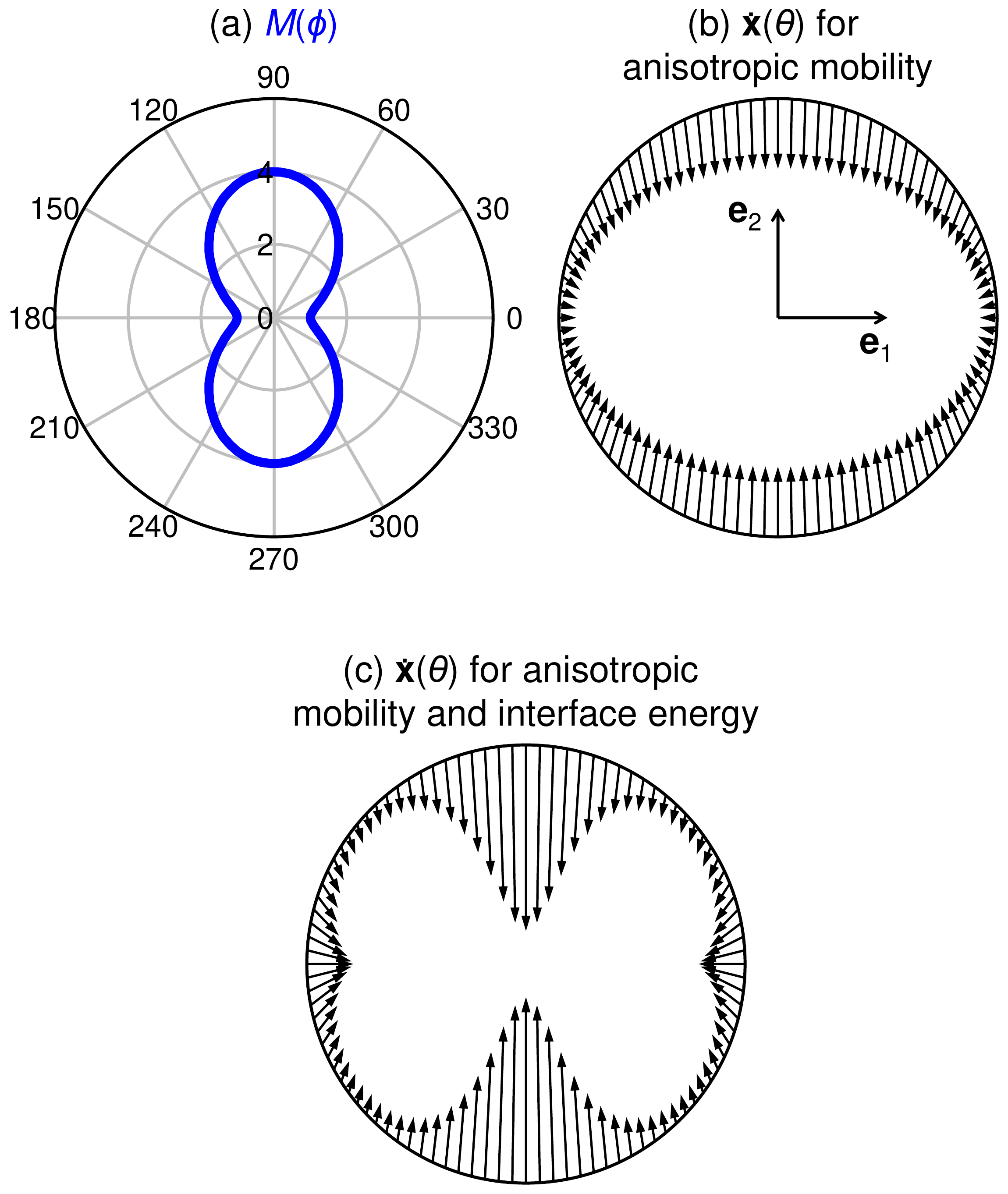}\hspace{-1.78em}%
\caption{\label{fig:anisotropykinetics}
(a) Polar plot of the interface mobility vs. inclination angle $M(\phi)$ (scaled by $M^{(1)}$) with $M^{(2)}/M^{(1)} = 4$. 
(b) Interface velocity  along a circular interface for isotropic interface energy. 
The kinetics are anisotropic; i.e., we set $M^{(2)}/M^{(1)} = 4$. 
(c) Interface velocity  in the case anisotropic interface energy (Eq.~\eqref{gammaaniso} with $\epsilon = 0.3$, $M^{(2)}/M^{(1)}=4$)  and $\gamma^{(1)} =\gamma^{(2)}$. 
}
\end{figure}

This implies that the interface velocity is 
\begin{equation}\label{kineticeq_s_M1M2}
\dot{\mathbf{x}}
= \Gamma \kappa \mathbf{M} \hat{\mathbf{n}}, 
\end{equation}
where $\mathbf{M}$ is the mobility tensor
\begin{equation}
\mathbf{M}
\equiv 
\left(\begin{array}{cc}
M^{(1)} & 0 \\ 
0 & M^{(2)}
\end{array}\right).
\end{equation}
The interface velocity $\dot{\mathbf{x}}(s)$ is parallel to $\mathbf{M}\hat{\mathbf{n}}$ rather than the interface normal  $\hat{\mathbf{n}}$. 
In the step-flow model of interface motion, the driving force acts on the interface steps, rather than the interface \emph{per se}.
When $M^{(1)} = M^{(2)} = M$, $\dot{\mathbf{x}} = M \Gamma \kappa \hat{\mathbf{n}}$, which is identical to the classical curvature flow equation of motion Eq.~\eqref{capillarity}. 
Equation~\eqref{kineticeq_s_M1M2} is a generalisation of Eq.~\eqref{capillarity} that accounts for kinetic anisotropy ($M^{(1)} \ne M^{(2)}$). 
{\color{black}
Note that, although Eq.~\eqref{kineticeq_s_M1M2} looks like the classical curvature flow equation, it is a consequence of the step-flow kinetics (rather than assumed). 
}
For a flat interface with a fixed inclination angle $\phi$, only the velocity projected along the interface normal contributes to interface migration (tangential components of the velocity are important for description of the motion of corners):
\begin{equation}
M(\phi)
= \hat{\mathbf{n}}\cdot\mathbf{M}\hat{\mathbf{n}}
= M^{(1)}\cos^2\phi + M^{(2)}\sin^2\phi, 
\end{equation}
as depicted in Fig.~\ref{fig:anisotropykinetics}a. 
This prediction is qualitatively similar  to the inclination dependence of grain boundary motion from kinetic Monte Carlo~\cite{lobkovsky2004grain} and molecular dynamics~\cite{lontine2018stress} simulations. 
We  simulate the interface evolution mediated by step migration of an arbitrarily curved interface by  numerically solving Eq.~\eqref{kineticeq_s_M1M2}.

If the interface energy is isotropic, $\gamma = \gamma_0$, then  the interface velocity is given by Eq.~\eqref{kineticeq_s_M1M2} with $\Gamma = \gamma_0$.
Figure~\ref{fig:anisotropykinetics}b shows an example of the velocity field on a circular interface (radius $R$) with $M^{(2)}/M^{(1)} = 4$. 
Here, the horizontal interface segment moves faster than the vertical interface segment because the horizontal interface segment is equivalent to R(2) steps while the vertical interface segment to R(1) steps and we set $M^{(2)} > M^{(1)}$ (i.e., Type~(2) migration is faster than Type~(1) migration, as seen in Fig.~\ref{fig:migrationmode}). 

For the anisotropic interface energy case, Eq.~\eqref{gammaaniso}, 
\begin{equation}\label{kinetic_anisotropy_regularisation}
\dot{\mathbf{x}}(s)
= \frac{2\epsilon}{\pi} 
\left(
\gamma^{(2)} \frac{\hat{l}_2^2}{\epsilon^2 + \hat{l}_1^2} 
+ \gamma^{(1)} \frac{\hat{l}_1^2}{\epsilon^2 + \hat{l}_2^2}
\right)
\kappa \mathbf{M} \hat{\mathbf{n}}.
\end{equation} 
Focusing on the velocity field around a circular interface $\dot{\mathbf{x}}(\theta)$, we see that the interface velocity (see Fig.~\ref{fig:anisotropykinetics}c for $\epsilon = 0.3$, $M^{(2)}/M^{(1)}=4$, and $\gamma^{(1)} =\gamma^{(2)}$) is larger on the horizontal segments ($\theta = \pi/2$ and $3\pi/2$) than on  vertical ones ($\theta = 0$ and $\pi$) since $M^{(2)} > M^{(1)}$; consistent with Fig.~\ref{fig:anisotropykinetics}b. 
When $M^{(1)} \ne M^{(2)}$, the interface shape does not evolve to the equilibrium shape (i.e., the lowest-energy profile at fixed grain/domain area).

\subsection{General disconnection}\label{disconnectionmodel}

As noted above disconnections generally have both step and dislocation characters (a pure-step model is more appropriate for describing the dynamics of free surfaces than interfaces in crystalline materials). 
Here, we present the two-reference general disconnection model. 

\begin{figure}[b!]
\includegraphics[height=0.5\linewidth]{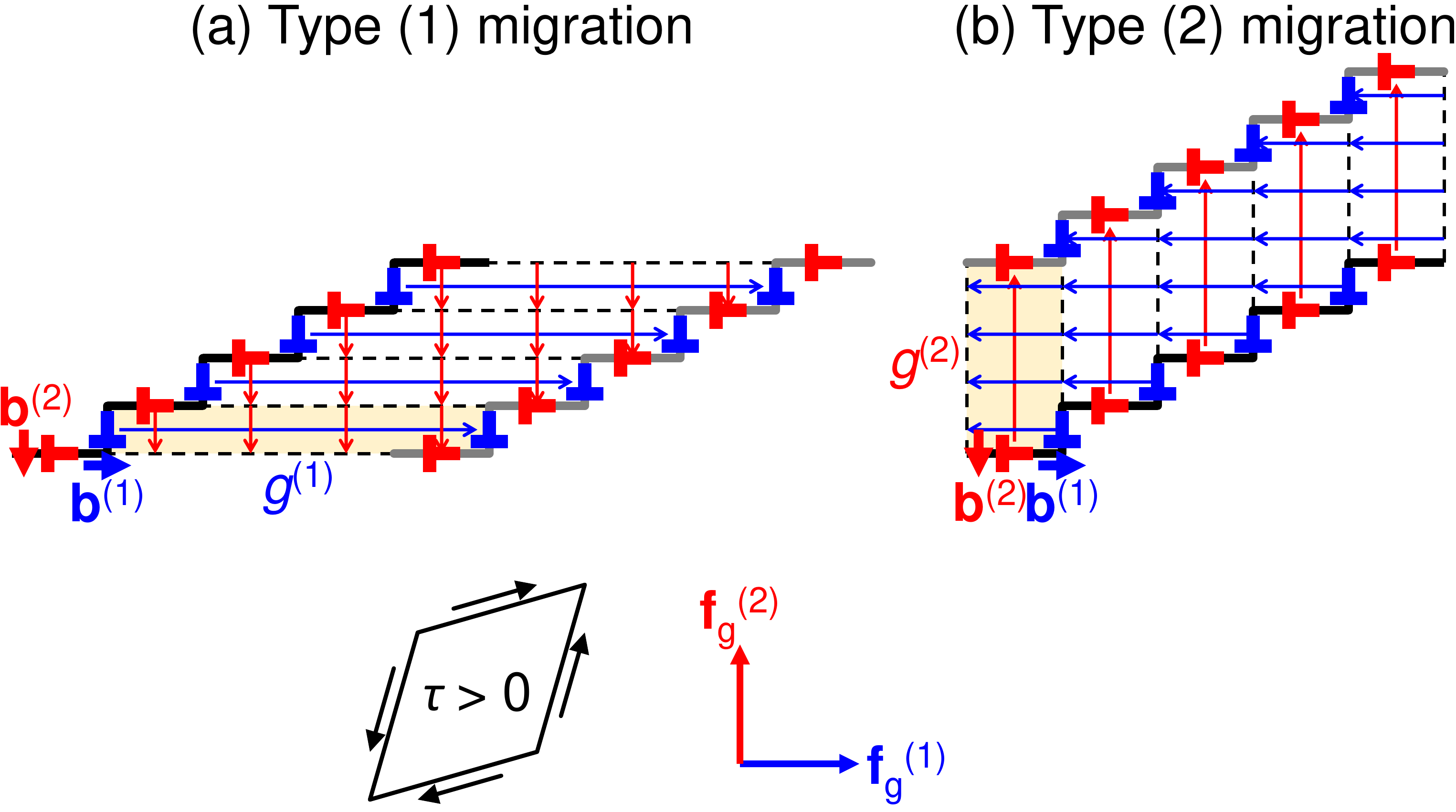}\hspace{-1.78em}%
\caption{\label{fig:migrationmode_b}Two  interface migration modes for disconnection  with finite Burgers vectors (cf. Fig.~\ref{fig:migrationmode} for the pure step case). 
This interface segment corresponds to Point A in Fig.~\ref{fig:tangentgeometry_b}. 
(a) Type~(1) migration is driven by  horizontal glide of R(1) disconnections. 
The shaded region shows that the glide of one R(1) disconnection towards the right is accompanied by the downward glide of R(2) disconnections. 
(b) Type~(2) migration is driven by the vertical glide of R(2) disconnections. 
The shaded region shows that the upward glide of an R( 2) disconnection is accompanied by the leftward glide of R(1) disconnections . 
 $g^{(1)}$ and  $g^{(2)}$ are glide distances and $f_\ug^{(k)}$ is the PK force on R($k$) disconnections under positive $\tau$. 
}
\end{figure}

We employ the following conventions: 
(i) the disconnection line direction $\boldsymbol{\xi} \parallel \mathbf{e}_3$, (ii) the Burgers vector is defined such that $\boldsymbol{\xi} \times \mathbf{b}$ points to the extra half plane, and (iii) $h^{(k)}\geq0$. 
The Peach-Koehler (PK) force is thus
\begin{equation}
\mathbf{f}_\text{PK} 
= (\boldsymbol{\upsigma}\cdot\mathbf{b}) \times \boldsymbol{\xi}
= (\boldsymbol{\upsigma}\cdot\mathbf{b}) \times \mathbf{e}_3, 
\end{equation}
where $\boldsymbol{\upsigma}$ is the total stress. 
The Burgers vector associated with an R($k$) disconnection is denoted by $\mathbf{b}^{(k)} = b^{(k)}\mathbf{e}_k$ ($k = 1,2$); $b^{(k)}$ can be positive or negative, depending on the bicrystallography (its magnitude and sign are fixed along the interface while the sign of $\rho^{(k)}$  varies along the interface). 
{\color{black}Here we explicitly assume that  disconnection glide kinetics are much faster than   disconnection climb. 
Our approach to selecting reference interfaces,  proposed in Sect.~\ref{theorytworef},  insures that we can always find short DSC lattice vectors parallel to the reference interface plane.
In other words, short glissile disconnections always exist as long as the reference planes are appropriately chosen.} 
For the disconnection case, the two types of pure-step interface migration (Fig.~\ref{fig:migrationmode}) are replaced by those shown in Fig.~\ref{fig:migrationmode_b}. 
Again, Type~(1) migration (see Fig.~\ref{fig:migrationmode_b}a) is driven by the glide of R(1) disconnections along the $\mathbf{e}_1$-axis (the R(2) disconnection motion is coupled) and  
Type~(2) migration (see Fig.~\ref{fig:migrationmode_b}b) is driven by the glide of R(2) disconnections along the $\mathbf{e}_2$-axis (coupled to the R(1) disconnection motion). 

The driving force on interface disconnections includes contributions from PK forces, capillarity, 
and chemical potential jumps 
{\color{black}(i.e.  jumps in the bulk free energy densities of two phases, or ``volume phase change part'' in Ref.~\onlinecite{taylor1992overview})}; we consider each separately. 
The R(1) disconnection Burgers vector density is 
\begin{equation}
\rho^{(1)} b^{(1)} = \frac{b^{(1)}}{h^{(1)}} h^{(1)}\rho^{(1)} = \beta^{(1)} \hat{l}_2, 
\end{equation}
where $\beta^{(k)} \equiv b^{(k)}/h^{(k)}$ is the shear-coupling factor~\cite{cahn2006coupling} for the R($k$) interface. 
{\color{black}The PK force projected onto the direction of motion of this set of disconnections} is $\tau \beta^{(1)} \hat{l}_2$ and the displacement is $\delta g^{(1)}$, where $\tau \equiv \sigma_{12}$ is the total shear stress on the interface. 
The energy variation with disconnection displacement is 
\begin{equation}
\delta E^{(1)}
= - \tau \beta^{(1)}\hat{l}_2 \delta g^{(1)}. 
\end{equation}

\begin{figure}[bt!]
\includegraphics[height=1.6\linewidth]{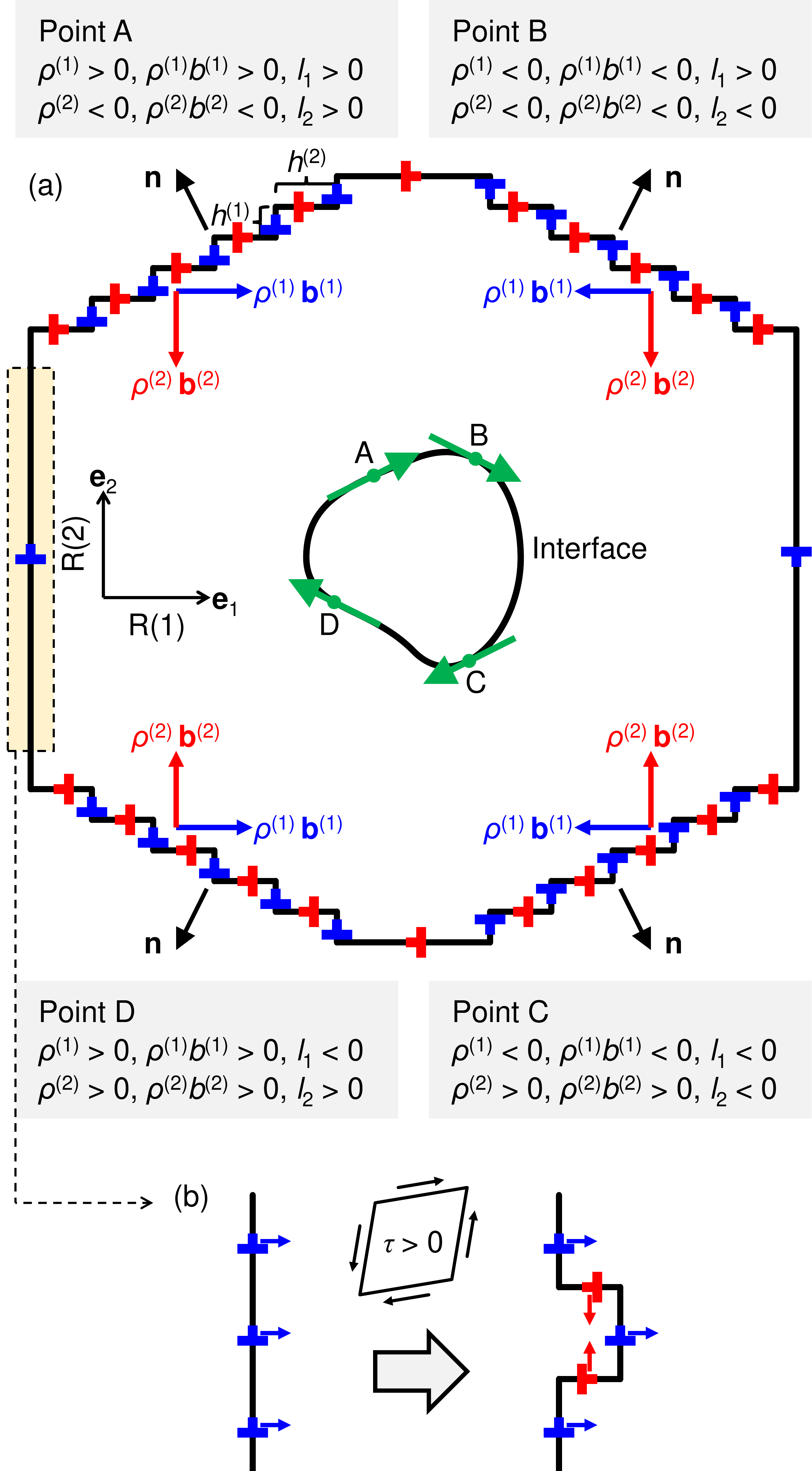}\hspace{-1.78em}%
\caption{\label{fig:tangentgeometry_b}
(a) A closed interface curve (centre) with the tangent vector oriented in the clockwise direction. 
Segments of length $|\mathbf{l}|$ are shown explicitly at Points A, B, C and D . 
The Burgers vector distributions associated with R(1) (blue) and R(2) (red) disconnections are shown for  $b^{(1)}> 0$ and $b^{(2)}> 0$ (The sign of the actual Burgers vector is determined by $\rho^{(k)} b^{(k)}$). 
(b) The motion of the interface segment between Points A and D under $\tau > 0$. 
The  arrows denote the PK force. 
}
\end{figure}

Considering Point A in Fig.~\ref{fig:tangentgeometry_b}a for $\tau > 0$ and $\delta g^{(1)} > 0$, we see that the energy decreases with R(1) disconnections $\mathbf{b}^{(1)}$ gliding to the right ($\delta E^{(1)} < 0$) and R(2) $\mathbf{b}^{(2)}$ gliding downwards. 
The glide of R(2) disconnections under $\tau$ also does work; 
\begin{equation}
\delta E^{(2)}_\text{coupling}
= \tau \beta^{(2)}
\hat{l}_2
\delta g^{(1)}. 
\end{equation}
Hence, the energy variation and the PK force associated with Type~(1) migration shown in Fig.~\ref{fig:migrationmode_b} are 
\begin{subequations}
\begin{gather}
\delta E
= \delta E^{(1)} + \delta E^{(2)}_\text{coupling}
= \tau \Lambda \hat{l}_2 \delta g^{(1)},
\end{gather}
\begin{gather}
f_\ug^{(1)}
= -\delta E / \delta g^{(1)}
= -\tau\Lambda\hat{l}_2, 
\end{gather}
\end{subequations}
where
\begin{equation}\label{defLambda}
\Lambda
\equiv \beta^{(2)} - \beta^{(1)}
= \frac{b^{(2)}}{h^{(2)}} - \frac{b^{(1)}}{h^{(1)}}.  
\end{equation}
Similarly, the PK force which drives Type~(2) migration (see Fig.~\ref{fig:migrationmode_b}b) is $f_\ug^{(2)} = -\delta E/\delta g^{(2)}= \tau\Lambda\hat{l}_1$. 
When $b^{(1)}/h^{(1)} = b^{(2)}/h^{(2)}$, $\Lambda = 0$ and $f_\ug^{(k)} = 0$. 
Consider the interface segment in Fig.~\ref{fig:tangentgeometry_b}b;  
positive $\tau$ drives R(1) disconnections to glide to the right, requiring nucleation and separation of an R(2) disconnection pair. 
But, nucleation and separation of the R(2) pair are inhibited by $\tau$. 
The net result is that this interface segment does not move upon application of stress $\tau$. 

Combining the capillary and stress effects and the two types of migration (Fig.~\ref{fig:migrationmode_b}) yields 
\begin{equation}\label{velocity_capillarity_stress_anisotropy}
\dot{\mathbf{x}}
= (\Gamma \kappa + \tau\Lambda) \mathbf{M}\hat{\mathbf{n}}. 
\end{equation}
As in the pure-step case, here the velocity $\dot{\mathbf{x}}$ is parallel to $\mathbf{M}\hat{\mathbf{n}}$ rather than the normal direction $\hat{\mathbf{n}}$. 
Comparing Eqs.~\eqref{kineticeq_s_M1M2}  and \eqref{velocity_capillarity_stress_anisotropy}, we see that, if the stress is dominated by a constant externally applied stress (as opposed to the stress fields of Burgers vectors on each other), the effect of stress is isotropic.  
We also see that, when $\beta^{(1)}=\beta^{(2)}$, 
$\Lambda = 0$ and $\tau$ does not affect interface motion.  

\begin{figure}[bt!]
\includegraphics[height=0.72\linewidth]{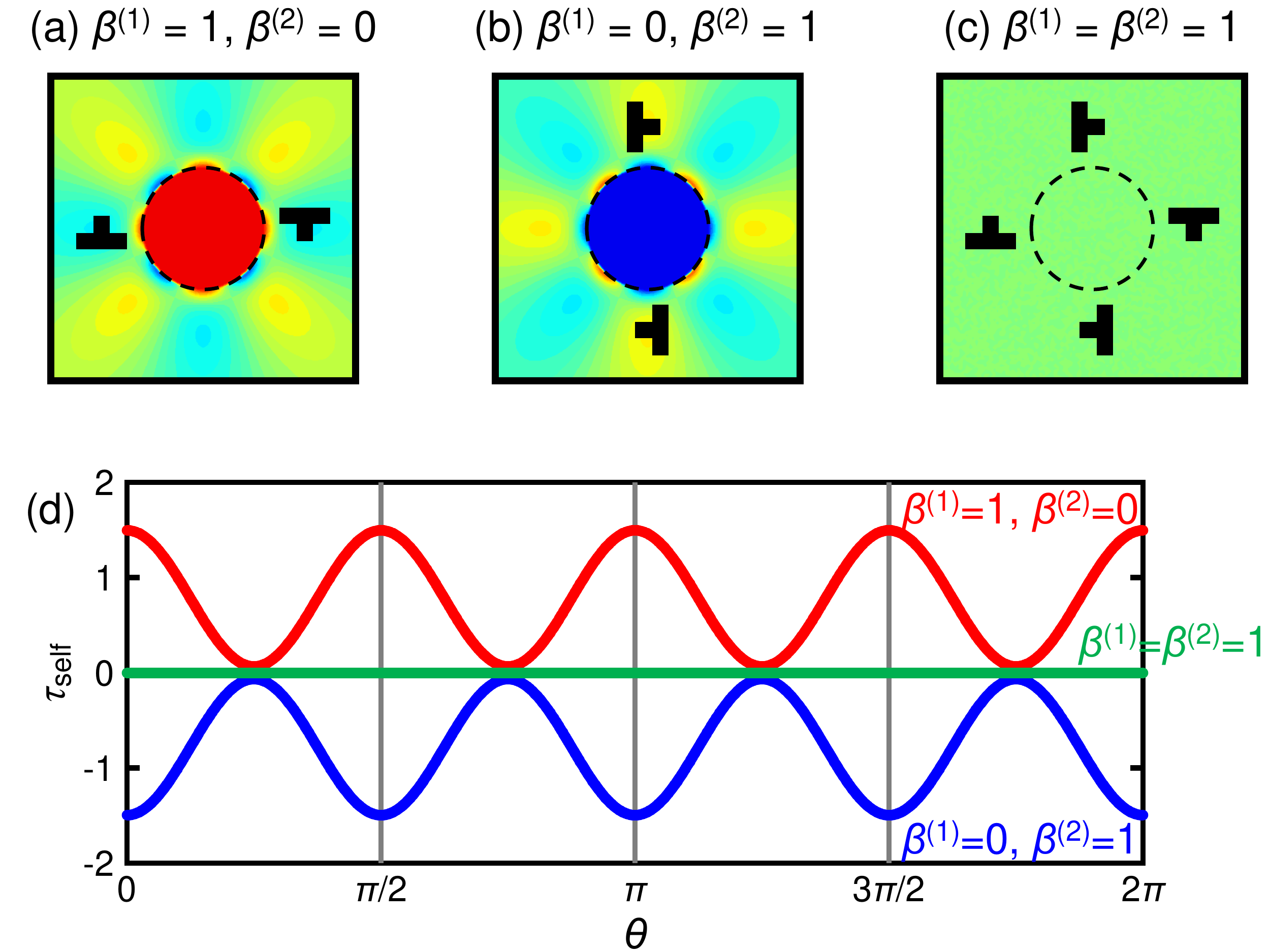}\hspace{-1.78em}%
\caption{\label{fig:s12_self_stress}
The stress fields $\tau_\text{self}(\mathbf{r})$ from the disconnections on a circular interface for (a) $\beta^{(1)} = 1$ and $\beta^{(2)} = 0$, (b) $\beta^{(1)} = 0$ and $\beta^{(2)} = 1$, and (c) $\beta^{(1)} = \beta^{(2)} = 1$. 
(d) The value of $\tau_\text{self}$ along the circular interface for the three cases. 
}
\end{figure}

Note that in the above treatment we ignore the elastic interaction among the disconnections along the interface. 
The internal (or \textit{self}) shear stress from all disconnections at each point along the interface can be written as
\begin{equation}\label{s12curve_s}
\tau_\text{self}(s)
= \beta^{(1)} I^{(1)}(s) + \beta^{(2)} I^{(2)}(s), 
\end{equation}
where
\begin{align}
I^{(m)}(s)
&= \frac{G}{2\pi(1-\nu)} \int
\left(\frac{\ud x_n}{\ud s}\right)_{s=s_0} 
\dfrac{x_m(s) - x_m(s_0)}{\varrho_a^2}
\nonumber\\
&\times \left[
1 - \dfrac{2 (x_n(s) - x_n(s_0))^2}{\varrho_a^2}
\right]
\ud s_0,
\end{align}
and $(m, n) = (1, 2)$ or $(2, 1)$, $\varrho_a^2 \equiv [x_1(s) - x_1(s_0)]^2 + [x_2(s) - x_2(s_0)]^2 + a^2$ ($a$ is the disconnection core size)~\cite{cai2006non}, $G$ is the shear modulus, and $\nu$ is the Poisson ratio.
These integrals must be evaluated  numerically. 
Figure~\ref{fig:s12_self_stress} shows these stresses for the case of a circular interface. 
Finally, the total shear stress should be $\tau = \tau_\text{self} + \tau_{\rm ext}$ with $\tau_{\rm ext}$ the external shear stress as considered above.

\subsection{Chemical potential jumps}

Diffusionless transformations (e.g., Martensite) often involve interface migration. 
Such phase transformations/interphase interface migration are induced by  differences  in chemical potentials between the two phases. 
A chemical potential jump across the interface between domains of different phases provides an additional, important driving force for interface migration. 

In the analysis of chemical potential jump effects, we label the interface normal  $\hat{\mathbf{n}}$ as shown in Fig.~\ref{fig:tangentgeometry_b} and use the notation ``$\pm$'' to denote the material property on the side of the interface to/from which the normal points.
The chemical potential jump is $\psi(s) = \mu^+(s) - \mu^-(s)$, where $\mu^{\pm}$ is the chemical potential of the phase on the ``${\pm}$'' side of the interface.

The glide of an R(1) disconnection (Fig.~\ref{fig:migrationmode_b}a) by $\delta g^{(1)}$ leads to an energy change $\psi h^{(1)} \delta g^{(1)}$. 
So, when an  R($k$) disconnection migrates  by  $\delta g^{(k)}$, the energy change is 
\begin{equation}
\delta E
= \psi \left(h^{(1)} \rho^{(1)} \delta g^{(1)} +  h^{(2)} \rho^{(2)} \delta g^{(2)}\right).
\end{equation}
The driving forces  on R($k$)  disconnections associated with the chemical potential jump are 
\begin{equation}\label{fcs}
f_\uc^{(1)} = - \frac{\delta E}{\delta g^{(1)}} = - \psi \hat{l}_2, \quad
f_\uc^{(2)} = - \frac{\delta E}{\delta g^{(2)}} = - \psi \hat{l}_1 .
\end{equation}
Based on the configuration shown in Fig.~\ref{fig:tangentgeometry_b}, if $\psi > 0$, the chemical potential in the matrix  is higher than in the embedded phase and the chemical potential jump favors the growth of the embedded phase. 
Combining this with the capillary and PK forces, the total driving forces along the interface are 
\begin{equation}
f^{(1)}
= - (\Gamma \kappa + \psi + \tau\Lambda) \hat{l}_2,
\quad
f^{(2)}
= (\Gamma \kappa + \psi + \tau\Lambda) \hat{l}_1. 
\end{equation}
such that the general kinetic equation for the motion of an  interface is
\begin{equation}\label{vMf}
\mathbf{v}
= \mathbf{M} \mathbf{f},
\end{equation}
where $\mathbf{M}$ is the mobility tensor and $\mathbf{f}$ is the driving force
\begin{equation}\label{fGptn}
\mathbf{f}
\equiv \left(\Gamma \kappa + \psi + \tau\Lambda\right) 
\hat{\mathbf{n}}. 
\end{equation}

\section{Numerical examples}\label{numerical}

We now examine the effects of the interface migration models presented above with the aid of numerical simulations. 
We consider here the  simple case of an initially circular domain embedded in a matrix (i.e., two phases or grains). 
The equation of motion for each model is solved via a finite difference approach.
More complex microstructures, with arbitrary domain morphologies and several interfaces with different properties (including also of topological change), are addressed in Part II of this paper through a diffuse interface approach~\cite{Salvalaglio2021}.  

\subsection{Reduced units and integration scheme}

We employ the following reduced units: $\tilde{\mathbf{x}} = \mathbf{x}/\alpha$ (for all ``length'' quantities),
$\tilde{\kappa} = \kappa \alpha$,
$\Delta\tilde{t} = \Delta t M^{(1)}\gamma^{(1)}/\alpha^2$, $\tilde{\gamma} = \gamma^{(2)}/\gamma^{(1)}$,  $\tilde{M} = M^{(2)} / M^{(1)}$, $\tilde{\tau} = \tau \alpha/\gamma^{(1)}$ and $\tilde{\psi} = \psi \alpha/\gamma^{(1)}$, we choose $\alpha$ as the length of one DSC cell edge. 

The curve $\tilde{\mathbf{x}}(s)$ representing the interface is discretised into $N$ nodes/segments, $\{ \tilde{\mathbf{x}}(i)\}$ with $n = 0, \cdots, N-1$. Derivatives $\tilde{\mathbf{x}}'$ and $\tilde{\mathbf{x}}''$ are computed using a three-point stencil $\{\tilde{\mathbf{x}}(i-1),\tilde{\mathbf{x}}(i),\tilde{\mathbf{x}}({i+1})\}$ and centred finite differences with periodic boundary conditions $\tilde{\mathbf{x}}(N)=\tilde{\mathbf{x}}(0)$ (for closed curves). 
All  other geometrical properties of the interface are computed from these quantities (e.g.,  $\kappa(i)$, $|\mathbf{l}(i)|$, $\hat{\mathbf{l}}(i)$ and $\hat{\mathbf{n}}(i)$).

After  initialisation of the node coordinates, $\{\tilde{\mathbf{x}}(i)\}$, the evolution is computed from the node velocities $(\tilde{v}_1(i), \tilde{v}_2(i))$ using  Eqs.~\eqref{vMf} and \eqref{fGptn} . 
We recall that these equations reduce to Eq.~\eqref{simplest} for pure steps and isotropic interface energy and mobility, to Eq.~\eqref{kinetic_anisotropy_regularisation} for pure steps and anisotropic interface energy and mobility, and to Eq.~\eqref{velocity_capillarity_stress_anisotropy} for disconnections with anisotropic properties but vanishing chemical potential difference across the interface. 
Moreover, to avoid node overlap, we connect each neighbouring node pair by a spring (spring constant $\mathcal{K}$), i.e., displace each node by an additional spring force along the tangent direction.  
Therefore, we consider the following integration scheme
\begin{equation}
\tilde{x}_n(i) 
:= 
\tilde{x}_n(i) 
+ \Delta\tilde{t}
\left[
\tilde{v}_n(i)
+ 
\tilde{\xi}(i)
\hat{l}_n(i)  
\right], 
\end{equation}
where $n=1,2$, $i$ is the index of node, and
\begin{equation}\label{tildexi}
\tilde{\xi}(i)
= \mathcal{K} \left[
|\tilde{\mathbf{x}}(i+1) - \tilde{\mathbf{x}}(i)|
- |\tilde{\mathbf{x}}(i) - \tilde{\mathbf{x}}(i-1)|
\right]. 
\end{equation}
In its general form (see Eq.~(32) and (33)), $\tilde{v}_n(i)$ reads
\begin{equation}
\bigg(
\def\arraystretch{1}
\begin{array}{c}
\tilde{v}_1(i)  \\
\tilde{v}_2(i) 
\end{array} \bigg)=
\left[
\tilde{\Gamma}(i) \tilde{\kappa}(i) 
+ \tilde{\psi} + \tilde{\tau}(i) \Lambda
\right]
\bigg(
\def\arraystretch{1}
\begin{array}{c}
-\hat{l}_2(i)  \\
\tilde{M} \hat{l}_1(i) 
\end{array} \bigg)
\end{equation}
with  interface stiffness
\begin{equation}\label{tildeGammai}
\tilde{\Gamma}(i)
= \frac{2\epsilon}{\pi}
\left[
\tilde{\gamma} \frac{\hat{l}_2^2(i)}{\epsilon^2 + \hat{l}_1^2(i)}
+ \frac{\hat{l}_1^2(i)}{\epsilon^2 + \hat{l}_2^2(i)}
\right], 
\end{equation}
and  total shear stress  
\begin{equation}
\tilde{\tau}(i)
= \tilde{G} \left[\beta^{(n)} \tilde{I}^{(1)}(i) + \beta^{(2)} \tilde{I}^{(2)}(i)\right]
+ \tilde{\tau}_\text{ext}, 
\end{equation}
where 
\begin{equation}
\begin{split}
&\tilde{I}^{(m)}(i)
= 
\sum_{j=0}^{N-1}
\tilde{x}_n'(j) \tilde{\tau}^{(m)}(i,j), \\
&\tilde{\tau}^{(m)}(i,j)
= \dfrac{\tilde{x}_m(i) - \tilde{x}_m(j)}{\tilde{\varrho}_a^2(i,j)}
\left\{
1 - \dfrac{2[\tilde{x}_n(i) - \tilde{x}_n(j)]^2}{\tilde{\varrho}_a^2(i,j)}
\right\},  
\\
&\tilde{\varrho}_a^2(i,j)
= [\tilde{x}_1(i) - \tilde{x}_1(j)]^2 + [\tilde{x}_2(i) - \tilde{x}_2(j)]^2 + \tilde{a}^2, 
\end{split}
\end{equation}
and $\tilde{G} = G \alpha /[2\pi(1-\nu) \gamma^{(1)}]$. 

For the  isotropic mobility case, we simply  set $\tilde{M}=1$, while isotropic interface energy implies $\tilde{\Gamma}(i)=\tilde{\gamma}_0 = 1$. 
Pure step motion can be recovered by setting $\beta^{(1)}=\beta^{(2)}=0$. Further details as well as the integration schemes corresponding to these different limits are presented in the Supplemental Material.

\subsection{Pure steps, isotropic energy and mobility}

\begin{figure}
\includegraphics[width=1\linewidth]{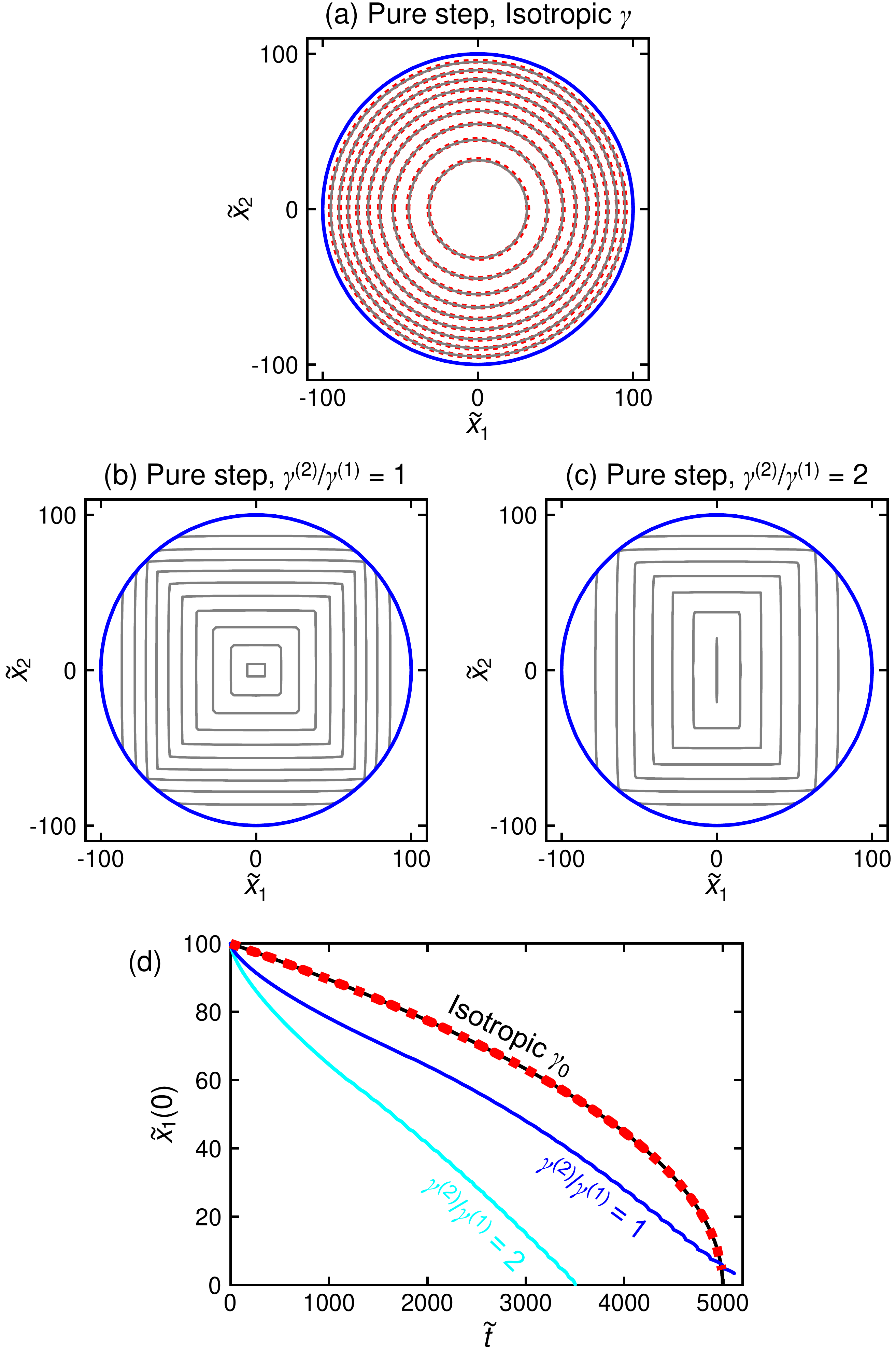}\hspace{-1.78em}%
\caption{\label{fig:PureStepEvolve}
(a)-(c) Evolution of an initially circular interface based on the pure-step model. 
The blue lines denote the initial interface profile and the grey lines show the simulation results at increments of $\Delta \tilde{t} = 500$. 
(a) Isotropic interface energy case (the red dotted curves show the analytical result). 
Anisotropic interface energies with (b) $\gamma^{(2)}/\gamma^{(1)} = 1$ and (c) $2$. 
(d) The position of the $0^\text{th}$node on the $\mathbf{e}_1$-axis $\tilde{x}_1(0)$ versus time $\tilde{t}$. 
$\tilde{x}_1(0)$ is a linear measure of the domain size. 
The black line and red dotted lines are the simulation result for the pure-step model and the analytical solution with isotropic interface energy.  
The blue and cyan lines are the simulation results for the pure-step model with anisotropic interface energy.
}
\end{figure}

To serve as a basis for evaluating  the effects of anisotropy and finite Burgers vector disconnections, we first examine the classical capillarity-driven interface motion in the isotropic interface energy and mobility and zero Burgers vector limit (i.e., Eq.~\eqref{simplest}).  
This also serves as a check on the fundamental two-reference approach presented above.
Initially ($\tilde{t} = 0$), the interface is a circle of radius $\tilde{R} = 100$ (the blue dashed curve in Fig.~\ref{fig:PureStepEvolve}a). 
The evolving profile is shown by a series of black curves (at different times) in Fig.~\ref{fig:PureStepEvolve}a. 
For this special case, the evolution of the initially circular profile is described by $\ud\tilde{R}/\ud\tilde{t} = - 1/\tilde{R}$.
The analytical solution to this equation is $\tilde{R} = (\tilde{R}_0^2 - 2\tilde{t})^{1/2}$. 
The embedded domain evolves with time as shown in Fig.~\ref{fig:PureStepEvolve}a or d, where the black and  dashed red curves show the simulation and analytical results, respectively. 
The consistency between the simulation and analytical results in Figs.~\ref{fig:PureStepEvolve}a and d validates the numerical implementation.

\subsection{Pure steps, anisotropic energy, isotropic mobility}

We now consider the case of interface migration via a two-reference, pure-step model with  anisotropic interface energy and isotropic mobility; i.e., Eq.~\eqref{capillarity_anisotropy_regularisation}. 
The interface energy is described by Eq.~\eqref{gammaaniso}, with parameters $\gamma^{(1)}$ and $\gamma^{(2)}$. 
The simulation results are shown in Figs.~\ref{fig:PureStepEvolve}b and c, where $\epsilon^2 = 10^{-4}$ and $\tilde{\gamma} =$ 1 or 2. 

Facets on the embedded domain begin to form immediately and spread during the evolution until the domain is completely faceted.
The facets correspond to the cusps in $\gamma(\phi)$ (e.g. Fig.~\ref{fig:anisotropy}a). 
At late time, the shape of the embedded domain resembles the  equilibrium shape of crystals obtained by the Wulff construction~\cite{balluffi2005kinetics}. 
When $\gamma^{(2)} > \gamma^{(1)}$, the R(2) steps disappear faster than the R(1) steps; this implies the R(2) interface flattens faster than the R(1) interface -- exactly as the results in Fig.~\ref{fig:anisotropy}b. 
Of course, when  $\gamma^{(1)} = \gamma^{(2)}$, the shape of the fully faceted and embedded domain is a square.

In Fig.~\ref{fig:PureStepEvolve}d, we plot the displacement of the $\theta = 0$  node as a function of time. 
The embedded domain  does not follow the classical parabolic law (isotropic case), but rather approaches a linear relation once the domain is fully faceted.  
This can be understood as follows. 
After the square or rectangular shape  is achieved, the curvatures of the four flat edges are nearly zero. 
The shrinkage rate is then controlled by the motion of the corners. 
The constant corner velocity can be described as controlled by disconnection nucleation at the corners. 
Here, the corner velocity depends on the magnitude of the regularisation parameter $\epsilon$. 
$\epsilon$ depends on the ratio of the junction mobility to the interface mobility (Sect.~\ref{discussion_singularities}). 
In either case, this suggests a linear shrinkage rate $\tilde{x}_1(0) \sim \tilde{t}$.

\subsection{Disconnections,  stress, isotropic capillarity}

\begin{figure}[bt!]
\includegraphics[width=1\linewidth]{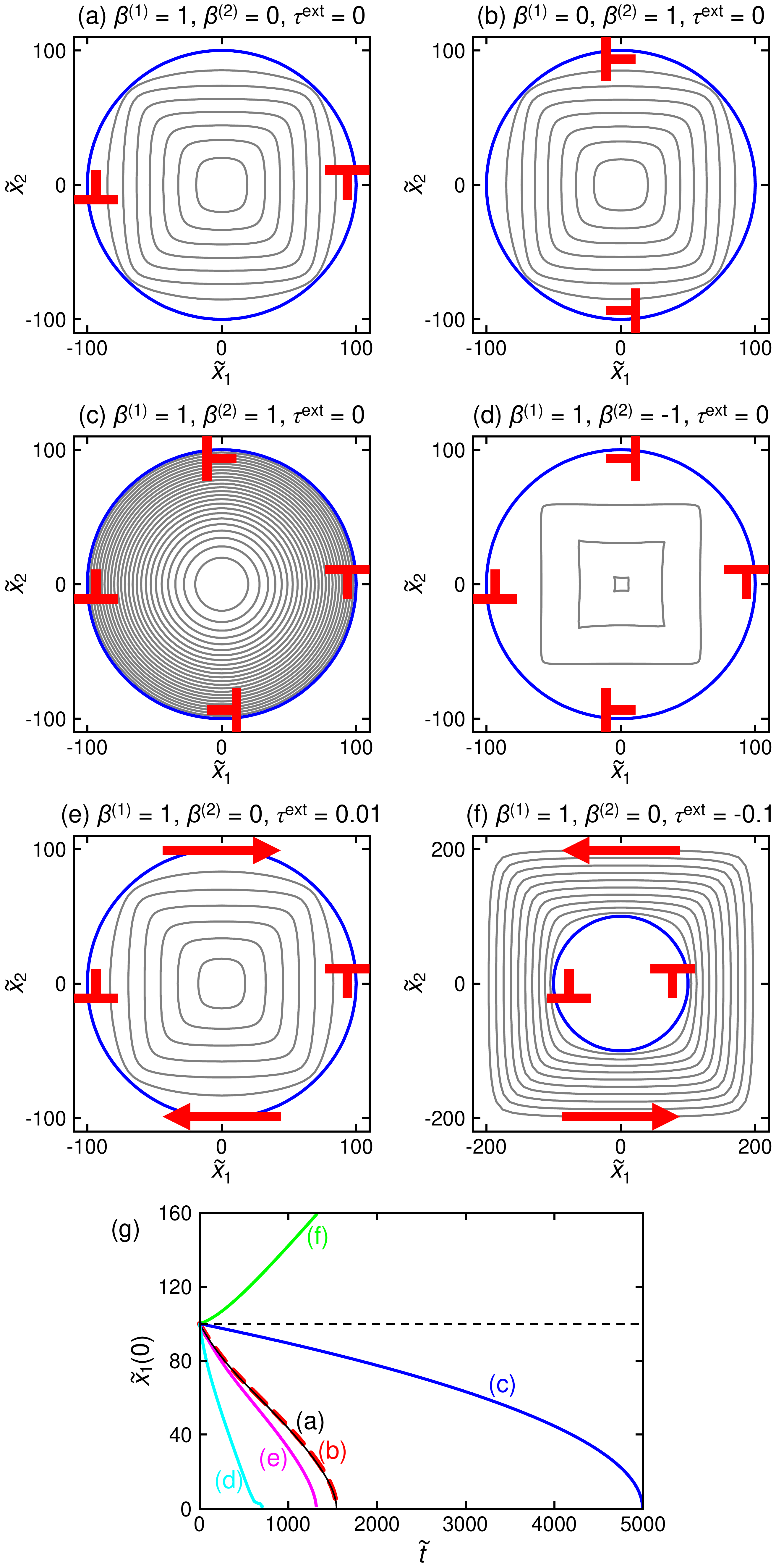}\hspace{-1.78em}%
\caption{\label{fig:OneModeEvolve}
(a)-(f) Evolution of an initially circular interface in which the disconnections have finite Burgers vector and the  interface energy is isotropic. 
The blue lines denote the interfaces at $\tilde{t} = 0$.
The time interval is $\Delta\tilde{t} = 200$. 
(g) The position of the $0^\text{th}$ node along the $\mathbf{e}_1$-axis $\tilde{x}_1(0)$ versus time $\tilde{t}$. 
$\tilde{x}_1(0)$ is a linear measure of the domain size. 
Each curve correspond to a case shown in (a)-(f). 
}
\end{figure}

Figure~\ref{fig:OneModeEvolve} shows simulation results for the evolution of the initially circular, embedded domain for disconnections with several different Burgers vectors and applied stresses. 
These simulations were performed based upon Eq.~\eqref{velocity_capillarity_stress_anisotropy}, where the   local stress $\tau(s)$ includes contributions from  all disconnections (Eq.~\eqref{s12curve_s}) and the applied stress and the interface energy and mobility are isotropic. 

Figure~\ref{fig:OneModeEvolve}a shows the case of $\beta^{(1)} = b^{(1)}/h^{(1)} =1$ and $\beta^{(2)} = 0$ (i.e., the R(2) disconnections are pure steps). 
The signs of the Burgers vectors on the left and right sides of the domain are opposite, such these two sides elastically attract one another. 
This attraction accelerates the shrinking of the embedded domain as compared with the pure-step, capillarity-driven case; cf. curve (a) in Fig.~\ref{fig:OneModeEvolve}a and the black curve in Fig.~\ref{fig:PureStepEvolve}d. 
We also note that there is a tendency towards faceting here even though the interface energy and mobility are isotropic. 
This faceting tendency originates from the elastic interactions between Burgers vectors. 
Figure~\ref{fig:OneModeEvolve}b shows the trajectory for the opposite case as that shown in Fig.~\ref{fig:OneModeEvolve}a; i.e., $\beta^{(1)} = 0$ and $\beta^{(2)} = 1$; the evolution of the two cases 
{\color{black}are (visually) indistinguishable}. 
Equation~\eqref{velocity_capillarity_stress_anisotropy} shows that the profile evolution  depends on $\tau\Lambda$;  
$\Lambda = -1$ and $1$ in the cases of Figs.~\ref{fig:OneModeEvolve}a and b. 
$\tau$ is also opposite between the Figs.~\ref{fig:OneModeEvolve}a and b cases. 
Hence, $\tau\Lambda$ is the same in both cases (as suggested by Eq.~\eqref{velocity_capillarity_stress_anisotropy}  and as seen here). 

Figure~\ref{fig:OneModeEvolve}c shows numerical results for $\beta^{(1)} = \beta^{(2)} = 1$. 
Here, $\Lambda = 0$ and hence  internal stress does not influence  profile evolution (implying that this case is identical to the pure-step, capillarity-driven migration case -- cv. curve (c) in Fig.~\ref{fig:OneModeEvolve}g and the black curve in Fig.~\ref{fig:PureStepEvolve}d). 
On the other hand, if we switch the sign of the Burgers vectors in the R(2) case (Fig.~\ref{fig:OneModeEvolve}d), $\Lambda = -2$ and the internal stress greatly accelerates the shrinkage of the embedded phase (curve (d) in Fig.~\ref{fig:OneModeEvolve}g). 
Again, the strong faceting is associated with elastic interactions rather than  anisotropic interface energy or mobility. 
The slightly negative curvature in the embedded domain walls at  late times is associated with the elastic attraction between  disconnections on  opposite sides. 

Figures~\ref{fig:OneModeEvolve}e and f show the same case as in Fig.~\ref{fig:OneModeEvolve}a but with an external/applied shear stresses $\tau_\text{ext}$ of opposite signs. 
In Fig.~\ref{fig:OneModeEvolve}e, the applied shear stress aids the annihilation of  disconnections of opposite sign on the left and right sides. 
In the positive $\tau_\text{ext}$ case (Fig.~\ref{fig:OneModeEvolve}e), the applied shear stress accelerates the shrinkage. 
However, in the negative $\tau_\text{ext}$ case (Fig.~\ref{fig:OneModeEvolve}f), the applied shear stress tends to separate the disconnections of opposite sign on the left and right sides. 
When the shear stress is sufficiently large, the PK force dominates the capillary force such that the embedded domain grows rather than shrinks. 
In this case, the embedded domain grows at a rate that approaches linear (see curve (f) in Fig.~\ref{fig:OneModeEvolve}g); this implies fundamentally different evolution kinetics/morphologies when the driving force is dominated by external stress or the elastic interactions between disconnections rather than by curvature. 

We do not simulate the case with finite chemical potential jump $\psi$ in Eq.~\eqref{fGptn}, since the effect of $\psi$ is trivial for a single interface. 
However, it becomes interesting and rather complex when we consider complex domain shapes or microstructures, as shown in Part II of this paper~\cite{Salvalaglio2021}.

\section{Discussion}\label{discussion}

\subsection{Reduction to the one-reference interface model}

We previously developed a one-reference, disconnection-based interface equation of motion~\cite{zhang2017equation} that assumes the interface deviates only slightly from a flat reference interface. 
We now consider the reduction of the present model to the one-reference limit in order to better understand its implications. 
We focus on the interface segment between Point A and Point B in Fig.~\ref{fig:tangentgeometry_b}. 
Take the R(1) interface as the only reference and assume that the AB interface segment deviates only slightly from this (i.e.,  $|\hat{l}_2/\hat{l}_1| \ll 1$ or, equivalently, $\hat{\mathbf{n}} \approx \mathbf{e}_2$).
The one-reference model also assumes that an interface migrates in the $\mathbf{e}_2$-direction via the lateral motion of disconnections; so, only Type~(1) migration (see Fig.~\ref{fig:migrationmode_b}a) occurs. 
To turn-off Type~(2) migration, we set $M^{(2)} = 0$ and $b^{(2)} = 0$. 
Then, from Eqs.~\eqref{vMf} and \eqref{fGptn}, 
\begin{equation}\label{v1l2}
\mathbf{v}
= v_1 \mathbf{e}_1
= -M^{(1)} \left(
\Gamma \kappa + \psi - \tau b^{(1)}/h^{(1)}
\right) \hat{l}_2 \mathbf{e}_1. 
\end{equation}
In the one-reference model, we can describe an interface curve by the height function $z = z(x_1)$ and assume that $|z_{,1}| \ll 1$; thus, $\kappa \approx z_{,11}$ and $\hat{l}_2 = z_{,1}/(1 + z_{,1}^2)^{1/2}$. 
The lateral velocity $v_1$ in Eq.~\eqref{v1l2} corresponds to the step velocity and the interface migration velocity is $\dot{z} = - v_1 \sgn(z_{,1}) (1 + z_{,1}^2)^{1/2}$. 
Thus, we obtain the equation of interface migration for a one-reference model:
\begin{equation}\label{Hdot}
\dot{z}
= M_\ud^{(1)} 
\left(\Gamma z_{,11} h^{(1)} + \psi h^{(1)} - \tau b^{(1)}\right)
\left(\left|z_{,1}\right| + B\right), 
\end{equation}
where $M_\ud^{(1)} \equiv M^{(1)} / h^{(1)}$ is the disconnection mobility. 
A brief explanation is required for the addition of the parameter $B$ in this expression. 
Without $B$, we find that $\dot{z} = 0$ when $z_{,1} = 0$; i.e., an infinitely large, flat R(1) interface will never migrate no matter how large the driving force is applied -- this does not make sense.
This issue was corrected by adding a term $B$ representing the thermal equilibrium concentration of disconnection dipoles~\cite{zhang2017equation}. 
So, for a flat R(1) interface, the kinetic equation is 
\begin{equation}\label{dotHone}
\dot{z}
= M_\ud^{(1)} B
\left(\psi h^{(1)} - \tau b^{(1)}\right).
\end{equation}
Equation~\eqref{Hdot} is that derived for grain boundary motion in a one-reference model~\cite{zhang2017equation}.

We  also revisit the  flat R(1) interface case based upon the two-reference model, without the assumptions inherent to the one-reference model. 
As before, we continue to assume that $b^{(2)}=0$. 
For a flat R(1) interface, we apply Eqs.~\eqref{vMf} and \eqref{fGptn} with $\hat{l}_1=1$ and $\hat{l}_2=0$:
\begin{equation}\label{dotHtwo}
\dot{z}
= M^{(2)}/h^{(1)} 
\left(\psi h^{(1)} - \tau b^{(1)}\right). 
\end{equation}
First, we see that the two-reference model does not require the addition of the new parameter $B$. 
Second, by comparing Eqs.~\eqref{dotHone} and \eqref{dotHtwo}, we have 
\begin{equation}
B = M^{(2)} / M^{(1)}. 
\end{equation}
Here, nucleation of disconnection dipoles involves Type~(2) migration. 
{\color{black}
We model the effect of disconnection nucleation/annihilation on R(1) interface migration by adjusting the ratio of the empirical ratio $M^{(2)}/M^{(1)}$. 
}

\subsection{Facet corners}\label{discussion_singularities}

It is not unusual for interface shapes to exhibits facets and/or sharp corners.
These may result from anisotropies in the interface energy/stiffness (e.g., cusps in the interface energy vs. interface normal) or in the interface mobility, as discussed above.
The resultant non-smoothness is commonly handled in computational studies and/or theories by some form of regularisation (e.g., Eq.~\eqref{capillarity_anisotropy}). 
Physically,  regularisation may be associated with atomic-scale discreteness, it is often invoked as a computational convenience on a scale much larger than atomic. 
In the regularisation approach discussed here (Eq.~\eqref{capillarity_anisotropy_regularisation}), this means identifying the physical origin of the regularisation parameter $\epsilon$. 
Alternatively, we can explicitly account for singularities in the interface profile. 

Consider the interface energy functional $E[\mathbf{x}] = \int F(s) \ud s$, where $F(s) \equiv \gamma(s) |\mathbf{l}(s)|$. 
If the interface shape is not smooth, $\mathbf{l}(s)$ is undefined at a  set of discrete points.
We divide the interface profile between singularities and  smooth sections, terminated by singularities; consider a smooth  interface section, terminating at  corners/junctions/singularities $s_1$ and $s_2$. 
The variation of the energy on this curved interface section  is
\begin{equation}\label{deltaEdeltaFint}
\delta E[\mathbf{x}] =
\left.\frac{\partial F}{\partial \mathbf{x}'}\cdot \delta\mathbf{x} \right|_{s_2^-} 
- \left.\frac{\partial F}{\partial \mathbf{x}'}\cdot \delta\mathbf{x} \right|_{s_1^+}
- \int \frac{\ud}{\ud s}\left(\frac{\partial F}{\partial \mathbf{x}'}\right) \cdot \delta\mathbf{x} \ud s, 
\end{equation}
where $s_1^+$ denotes approaching $s_1$ from $s>s_1$ and $s_2^-$ denotes approaching $s_2$ from $s<s_2$. 
{\color{black}
In Eq.~\eqref{deltaEdeltaFint} we evaluate the one-sided derivative of $F$; i.e., $F$ need not be differentiable with respect to $\mathbf{x}'$. 
}
At all points on the interface other than $s_1$ and $s_2$,  there is a capillarity driving force of the form of Eq.~\eqref{capillarity}. 

The driving forces on the singularities $s_1$ and $s_2$ are, respectively,  
\begin{equation}
\mathbf{f}_{s_1^+}
= \left.\frac{\partial F}{\partial \mathbf{x}'}\right|_{s_1^+}
\quad\text{and}\quad
\mathbf{f}_{s_2^-}
= -\left.\frac{\partial F}{\partial \mathbf{x}'}\right|_{s_2^-}. 
\end{equation}
Smooth interface sections meet at corners/singularities such that the total driving force on singularity $s$ is 
\begin{equation}
\mathbf{f}_{s}
= \left.\frac{\partial F}{\partial \mathbf{x}'}\right|_{s+}
-\left.\frac{\partial F}{\partial \mathbf{x}'}\right|_{s-}. 
\end{equation}
Overdamped  corner/singularity motion implies 
\begin{equation}\label{junctionmotion}
\mathbf{v}_s
= \mathscr{M} \mathbf{f}_s, 
\end{equation}
where $\mathscr{M}$ is the  mobility of the singularity point. 
Note, that $\mathscr{M}$ has different dimensionality than the interface mobility $M$ (see Eq.~\eqref{simplest} or \eqref{capillarity_anisotropy}); i.e., $M$ has dimension force/area while $\mathscr{M}$ has dimension force/length. 
We suggest that $\mathscr{M}$ may be viewed as a physical parameter, distinct from $M$ because the atomic-scale structure of a junction differs  from the interface itself.

How are the regularisation scheme (Eq.~\eqref{capillarity_anisotropy_regularisation}) and the discrete treatment of singularities related? 
Consider the special case of a square domain and focus on the velocity of the upper right corner of the square interface, for the interface energy of Eq.~\eqref{gammaaniso} with  $\gamma^{(1)} = \gamma^{(2)}$. 
When regularisation (Eq.~\eqref{capillarity_anisotropy}) is applied, sharp singularities (corner) are replaced with corners with a small curvature radius.
If the interface is of thickness $r_0$, a reasonable choice of corner curvature is $1/r_0$ and the corner velocity (Eq.~\eqref{capillarity_anisotropy}) is
$\mathbf{v}_s \approx -(4\epsilon M / \sqrt{2}\pi r_0) \gamma^{(1)}(1, 1)^T$ with $\epsilon \ll 1$. 
Equating this velocity with that from Eq.~\eqref{junctionmotion} ($\mathbf{v}_s = -\mathscr{M}\gamma^{(1)} (1, 1)^T$) yields
\begin{equation}
\epsilon
\sim r_0\mathscr{M} / M. 
\end{equation}
This relationship allows us to interpret the regularisation parameter $\epsilon$ as the ratio of the junction mobility and the interface mobility normalised by the interface thickness. 
In many materials, interfaces tend to be faceted (flat interfaces and sharp corners) at low temperatures, but smoother at high temperatures.  
This implies that at low temperature, $r_0\mathscr{M} \ll M$ ($\epsilon$ is very small), while at high temperature, $r_0\mathscr{M} / M \to 1$ ($\epsilon$ is  larger).

\subsection{Multi-mode ansatz}

Robust motion of  crystalline interfaces generally requires the existence and operation of multiple disconnection modes, associated with different Burgers vectors and step heights (there are an infinite set of possibilities for any interface)~\cite{han2018grain}.
At finite temperature, multiple modes may be activated at different rates such that the dynamics of any interface is a  statistical average of the contribution from  all active disconnection modes~\cite{chen2020grain}. 
We  previously included  multiple disconnection modes in a one-reference interface model to predict the temperature dependance of grain boundary motion~\cite{wei2019continuum}. 
How can multi-mode effects be incorporated in the two-reference model to describe general interface motion?

A multi-mode ansatz for the two-reference model replaces the shear-coupling factors, $\beta^{(1)}$ and $\beta^{(2)}$ with  {\it effective shear-coupling factors}, $\bar{\beta}^{(1)}$ and $\bar{\beta}^{(2)}$. 
Phenomenologically,  the effective shear-coupling factor $\bar{\beta}^{(k)}$ is  the ratio of the average interface sliding velocity (tangential velocity of the two crystals meeting at the interface) to the average interface (normal) velocity~\cite{han2018grain,chen2019grain}.
Molecular dynamics simulations~\cite{chen2019grain} of flat grain boundaries demonstrate that multiple disconnection modes show that the effective shear-coupling factor is a function of the local stress, chemical potential jump and temperature:
\begin{equation}
\bar{\beta}^{(k)}(\tau, \psi, T)
= \dfrac{\displaystyle{ \sum_m b^{(k,m)} e^{-\frac{Q^{(k,m)}}{k_\uB T}} \left(b^{(k,m)}\tau + h^{(k,m)}\psi\right) }}
{\displaystyle{ \sum_m h^{(k,m)} e^{-\frac{Q^{(k,m)}}{k_\uB T}} \left(b^{(k,m)}\tau + h^{(k,m)}\psi\right) }}, 
\end{equation}
where $b^{(k,m)}$, $h^{(k,m)}$ and $Q^{(k,m)}$ are the Burgers vector,  step height and  activation energy for disconnection mode $m$ on the R($k$) interface~\cite{chen2019grain}. 


\subsection{Non-orthogonal references}
\label{sec:non-orth}

In our presentation above, we focused on orthogonal reference interfaces; this is not always appropriate.  
Additionally, there are several reasonable reference interface choices (i.e.,  highly coherent interfaces associated with close-packed CSL planes). 
The present model can be generalised to account for these.

\begin{figure}[ht!]
\includegraphics[height=0.44\linewidth]{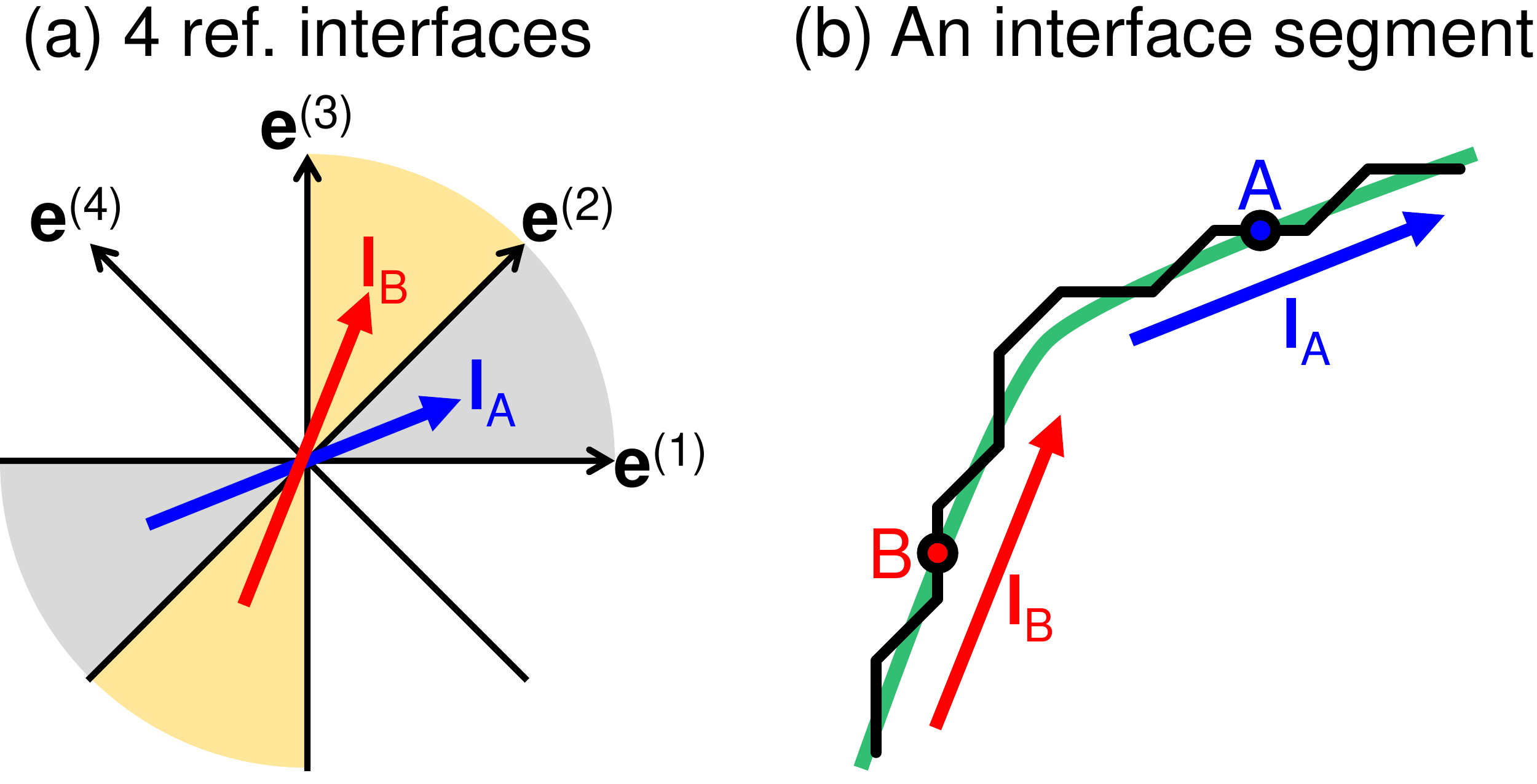}\hspace{-1.78em}%
\caption{\label{fig:nonorthogonal}
Interface  \textbf{l} and reference interface  \textbf{e}  orientations.
(a) $\{\mathbf{e}^{(k)}\}$ ($k=1, 2, 3, 4$) are four reference interfaces  tangents. 
(b) The green curve denotes an interface segment. 
The tangent vector at Point A (B) is $\mathbf{l}_\uA$ ($\mathbf{l}_\uB$). 
Since $\mathbf{l}_\uA$ ($\mathbf{l}_\uB$) lies in the grey (yellow) region delimited by $\mathbf{e}^{(1)}$ and $\mathbf{e}^{(2)}$ ($\mathbf{e}^{(2)}$ and $\mathbf{e}^{(3)}$) in (a), the interface at Point A can be described based on R(1) and R(2) (R(2) and R(3)) interfaces. 
}
\end{figure}

The choice of the two reference interfaces depends on the local interface tangent vector. 
As illustrated in Fig.~\ref{fig:nonorthogonal}a, we may assume that there are several potential reference interfaces $\{\mathbf{e}^{(k)}\}$ (e.g., $k=1,2,3,4$) that need not be orthogonal to one another. 
Denoting the Cartesian coordinate axes as $\mathbf{e}_1$ and $\mathbf{e}_2$, the inclination angle of each reference interface may be written: ${\phi^{(k)}}=\arcsin(\mathbf{e}^{(k)}\cdot \mathbf{e}_1)$. 
At a point $s$ on the interface has  local inclination  ${\phi^{(k)}}(s)=\arcsin(\mathbf{l}(s)\cdot \mathbf{e}_1)$ where $\mathbf{l}(s)$ is the local tangent. 
If $\phi(s)\in[\phi^{(k)},\phi^{(k+1)}]$, select reference interfaces R($k$) and R($k+1$). 
For example,  the orientation $\mathbf{l}_\uA$ in   Fig.~\ref{fig:nonorthogonal} is delimited by $\mathbf{e}^{(1)}$ and $\mathbf{e}^{(2)}$, such that the appropriate reference interfaces for Point A are R(1) and R(2). 
If R($k$) and R($k+1$) interfaces are determined for each point along the interface curve, we repeat the analyses presented above for these reference interfaces using the bases $\mathbf{e}^{(k)}$ and $\mathbf{e}^{(k+1)}$   for all vectors and tensors rather than $\mathbf{e}_1$ and $\mathbf{e}_2$. 
Such an analysis leads to a generalisation of Eqs.~\eqref{vMf} and \eqref{fGptn}, namely 
\begin{equation}\label{eq:nonorth1}
\mathbf{v} = \mathbf{M} \mathbf{f}, \qquad \mathbf{f}
\equiv \left(\Gamma\kappa + \psi + \boldsymbol{\tau} \cdot \boldsymbol{\lambda}\right) \hat{\mathbf{n}},
\end{equation}
for reference interfaces $\{\mathbf{e}^{(k)}, \mathbf{e}^{(k+1)}\}$. 
The mobility tensor in Eq.~\eqref{eq:nonorth1} is 
\begin{equation}
\mathbf{M}=M^{(k)}\mathbf{e}^{(k)}\otimes\mathbf{e}^{(k)}+M^{(k+1)}\mathbf{e}^{(k+1)}\otimes\mathbf{e}^{(k+1)} 
\end{equation}
and the anisotropic interface energy relative to  R($k$) and R($k+1$) is
\begin{equation}\label{eq:nonorth2}
\gamma=
\gamma^{(k+1)} \left|
\dfrac{\sin(\phi^{(k+1)} - \phi)}{\sin(\phi^{(k+1)} - \phi^{(k)})}
\right|
+
\gamma^{(k)} \left|
\dfrac{\sin(\phi - \phi^{(k)})}{\sin(\phi^{(k+1)} - \phi^{(k)})}
\right|,
\end{equation}
(the interface stiffness is $\Gamma = \gamma + \gamma_{,\phi\phi}$). 
Both the interface energy and mobility then reflect the symmetry of the reference interfaces. 
For the system illustrated in Fig.~\ref{fig:nonorthogonal}, there are eight minima in $\gamma(\phi)$ and up to eight maxima in $M(\phi)$ -- consistent with experimental and simulation results~\cite{kirch2008inclination,lontine2018stress}.
In Eq.~\eqref{eq:nonorth1}, $\boldsymbol{\tau}$ is the shear stress resolved on the reference interfaces $(\tau^{(k)}, \tau^{(k+1)})$, where
\begin{equation}\label{eq:nonorth8}
\tau^{(k)}
= \frac{\sigma_{22} - \sigma_{11}}{2} \sin 2\phi^{(k)}
+ \sigma_{12} \cos 2\phi^{(k)}.
\end{equation}
$\boldsymbol{\lambda}$ is the shear-coupling factor $(-\beta^{(1)}, \beta^{(2)})^T$. 
See the Supplemental Material for more details.
 

\section{Conclusions}

We present a comprehensive two-reference, interface model and equation of motion for arbitrarily inclined, curved interface that respects the underlying crystallographic relationships between  adjacent crystalline grains/phases. 
Our model incorporates the underlying disconnection-based interface migration mechanisms on each reference interface. 
This equation of motion also accounts for anisotropic interface energies and   mobilities and a wide range of driving forces -- capillarity, chemical potential difference and shear stress.

We also showed numerical results for the evolution of simple domain shapes in a  sharp interface model.
These results show interface faceting associated with thermodynamic  (anisotropic interface energy) and dynamic  (anisotropic interface mobility) effects and how these leads to very large differences with classical evolution laws (for capillarity- and stress-driven shape evolution). 
We also demonstrated how  disconnection character influences interface migration rates and morphology evolution.

Based on the model and theory presented here, we also developed a diffuse interface simulation approach appropriate for large-scale simulation of microstructure evolution and complex domain morphology evolution.  This is presented in Part II of this paper~\cite{Salvalaglio2021}.
{\color{black}
Generalisation to three dimensions follows naturally from the present model. Such an implementation, while complex in the largely analytical framework present here, is more naturally accommodated through  the diffuse interface approach developed in Part II. 
}

\section*{Acknowledgements}
JH acknowledges support from City University of Hong Kong Start-up Grant 7200667 and Strategic Research Grant (SRG-Fd) 7005466. 
DJS gratefully acknowledges the support of the  Hong Kong Research Grants Council Collaborative Research Fund C1005-19G. 
MS acknowledges support from Visiting Junior Fellowship of the Hong Kong Institute for Advanced Studies and the Emmy Noether Programme of the German Research Foundation (DFG) under Grant SA4032/2-1.

\providecommand{\noopsort}[1]{}\providecommand{\singleletter}[1]{#1}%
%

{\textcolor{white}{
\widetext
}}
\newpage
\onecolumngrid
\begin{center}
\textbf{\large {\sc{supplementary material}}\\
Disconnection-Mediated Migration of Interfaces in Microstructures:\\
I. continuum model}\\[.3cm]
Jian Han,$^{1}$ David J. Srolovitz$^{2}$, Marco Salvalaglio$^{3,4,5}$\\[.1cm]
  {\itshape \small
  ${}^1$Department of Materials Science and Engineering, City University of Hong Kong, Hong Kong SAR, China\\
    ${}^2$Department of Mechanical Engineering, The University of Hong Kong, Pokfulam Road, Hong Kong SAR, China\\
  ${}^3$Hong Kong Institute for Advanced Study, City University of Hong Kong, Hong Kong SAR, China\\
  ${}^4$Institute  of Scientific Computing,  TU  Dresden,  01062  Dresden,  Germany,
  \\ 
  ${}^5$Dresden Center for Computational Materials Science (DCMS), TU  Dresden,  01062  Dresden,  Germany}
\end{center}
\vspace{10pt}
\setcounter{equation}{0}
\setcounter{figure}{0}
\setcounter{table}{0}
\setcounter{section}{0}
\setcounter{page}{1}
\makeatletter
\renewcommand{\thesection}{S\Roman{section}}
\renewcommand{\theequation}{S\arabic{equation}}
\renewcommand{\thefigure}{S\arabic{figure}}
\renewcommand{\bibnumfmt}[1]{[S#1]}
\renewcommand{\citenumfont}[1]{S#1}

\twocolumngrid

\section{Numerical implementation}

\subsection{Finite difference}

Parameterise the interface profile as $\mathbf{x}(s)$ and discretise it as a  chain of $N$ nodes: $\{\mathbf{x}(i)\}$ ($i = 0, \cdots, N-1$) (we  choose the node indices to be the parameter $s$ and set the interval to be $\Delta s = 1$). 

First, we approximate the first derivative $\mathbf{l} = \mathbf{x}' = \ud\mathbf{x}/\ud s$ at each node. 
At  node $s_0$, 
\begin{equation}\label{expansion_Ds}
\left(\frac{\ud\mathbf{x}}{\ud s}\right)_{s_0}
= \frac{\mathbf{x}(s_0 + \Delta s) - \mathbf{x}(s_0 - \Delta s)}{2\Delta s}
+ \mathcal{O}[(\Delta s)^3]
\end{equation}
or as a finite difference, 
\begin{equation}\label{FD_D}
\mathbf{l}(i)
= \mathbf{x}'(i)
= [\mathbf{x}(i+1) - \mathbf{x}(i-1)]/2.  
\end{equation}
The second derivative $\mathbf{x}'' = \ud^2\mathbf{x}/\ud s^2$ at each node. 
At node $s_0$, 
\begin{equation}
\left(\frac{\ud^2\mathbf{x}}{\ud s^2}\right)_{s_0} 
= \frac{\mathbf{x}(s_0 + \Delta s) + \mathbf{x}(s_0 - \Delta s) - 2\mathbf{x}(s_0)}{(\Delta s)^2}
+ \mathcal{O}[(\Delta s)^4]
\end{equation}
or as a finite difference, 
\begin{equation}\label{FD_DD}
\mathbf{x}''(i)
= \mathbf{x}(i+1) + \mathbf{x}(i-1) - 2\mathbf{x}(i). 
\end{equation}
All  other quantities (on each node) may be easily evaluated based on $\mathbf{x}'(i)$ in Eq.~\eqref{FD_D} and $\mathbf{x}''(i)$ in Eq.~\eqref{FD_DD} by applying simple geometric definitions at node $i$: 
\begin{itemize}
\item[(i)]
The curvature  is 
\begin{equation}
\kappa(i)
= \dfrac{x_1'(i) x_2''(i) - x_1''(i) x_2'(i)}{\left[x_1'^2(i) + x_2'^2(i)\right]^{3/2}}.
\end{equation}

\item[(ii)]
The arc length (for $\ud s = 1$) is 
\begin{equation}
|\mathbf{l}(i)| = |\mathbf{x}'(i)|
= \sqrt{x_1'^2(i) + x_2'^2(i)}.
\end{equation}

\item[(iii)]
The unit tangent vector is 
\begin{equation}
\hat{\mathbf{l}}(i)
= \left(\begin{array}{c}
\hat{l}_1(i) \\ \hat{l}_2(i)
\end{array}\right)
= \frac{1}{|\mathbf{l}(i)|}\left(\begin{array}{c}
x_1'(i) \\ x_2'(i)
\end{array}\right).
\end{equation}

\item[(iv)]
The unit normal vector is
\begin{equation}
\hat{\mathbf{n}}(i)
= \left(\begin{array}{c}
-\hat{l}_2(i) \\ \hat{l}_1(i)
\end{array}\right)
= \frac{1}{|\mathbf{l}(i)|}\left(\begin{array}{c}
-x_2'(i) \\ x_1'(i)
\end{array}\right). 
\end{equation}
\end{itemize}

\subsection{Pure steps, isotropic energy and mobility}
\label{appendix2}

An algorithm for evolving a chain according to  kinetic equation Eq.~(8) at node $i$ is
\begin{equation}
\left(
\def\arraystretch{1}
\begin{array}{c}
\tilde{v}_1(i)  \\
\tilde{v}_2(i) 
\end{array}\right)
= \tilde{\kappa}(i)
\left(
\def\arraystretch{1}
\begin{array}{c}
-\hat{l}_2(i)  \\
\hat{l}_1(i) 
\end{array}\right). 
\end{equation}
This algorithm does not work in practice. 
When a node is very close to one of its neighbours, the curvature  at this node is nearly zero no matter where is the other neighbour -- this is unreasonable.
Hence, to avoid the overlap of two neighbouring nodes, we connect each neighbour node pair by a spring with spring constant $\mathcal{K}$, i.e., displace each node by the spring force along the tangent direction as per the following algorithm:
\begin{itemize}
\item[(i)]
Initialise the coordinates of all $N$ nodes: $\{\tilde{\mathbf{x}}(i)\}$ ($i = 0, \cdots, N-1$). 

\item[(ii)]
Displace each node by 
\begin{align}
&\left(
\def\arraystretch{1}
\begin{array}{c}
\tilde{x}_1(i)  \\
\tilde{x}_2(i) 
\end{array}\right)
:= 
\left(
\def\arraystretch{1}
\begin{array}{c}
\tilde{x}_1(i)  \\
\tilde{x}_2(i) 
\end{array}\right)
\nonumber\\
&+ \Delta\tilde{t}
\left[
\tilde{\kappa}(i)
\left(
\def\arraystretch{1}
\begin{array}{c}
-\hat{l}_2(i)  \\
\hat{l}_1(i) 
\end{array}\right)
+ 
\tilde{\xi}(i)
\left(
\def\arraystretch{1}
\begin{array}{c}
\hat{l}_1(i)  \\
\hat{l}_2(i) 
\end{array}\right)
\right], 
\end{align}
where
\begin{equation}\label{tildexi}
\tilde{\xi}(i)
= \mathcal{K} \left[
|\tilde{\mathbf{x}}(i+1) - \tilde{\mathbf{x}}(i)|
- |\tilde{\mathbf{x}}(i) - \tilde{\mathbf{x}}(i-1)|
\right]. 
\end{equation}
\end{itemize}
Since this force is tangential,  nodal separation is maintained, while not affecting motion along the normal.

\subsection{Pure steps, anisotropic energy and mobility}\label{Purestepsanisotropicenergy}

We employ the following algorithm for evolving a chain of nodes by kinetic equation Eq.~(19). 
First, define reduced quantities as: 
$\Delta\tilde{t} = \Delta t M^{(1)} \gamma^{(1)}/\alpha^2$, 
$\tilde{\gamma} = \gamma^{(2)}/\gamma^{(1)}$, 
and $\tilde{M} = M^{(2)} / M^{(1)}$. 
\begin{itemize}
\item[(i)]
Initialise the coordinates of all $N$ nodes: $\{\tilde{\mathbf{x}}(i)\}$ ($i = 0, \cdots, N-1$). 

\item[(ii)]
Displace each node by 
\begin{align}
&\left(
\def\arraystretch{1}
\begin{array}{c}
\tilde{x}_1(i)  \\
\tilde{x}_2(i) 
\end{array}\right)
:= 
\left(
\def\arraystretch{1}
\begin{array}{c}
\tilde{x}_1(i)  \\
\tilde{x}_2(i) 
\end{array}\right)
\nonumber\\
&+ \Delta\tilde{t} 
\left[
\tilde{\Gamma}(i) \tilde{\kappa}(i) 
\left(
\def\arraystretch{1}
\begin{array}{c}
-\hat{l}_2(i)  \\
\tilde{M} \hat{l}_1(i) 
\end{array}\right)
+ 
\tilde{\xi}(i)
\left(
\def\arraystretch{1}
\begin{array}{c}
\hat{l}_1(i)  \\
\hat{l}_2(i) 
\end{array}\right)
\right],
\end{align}
where
\begin{equation}\label{tildeGammai}
\tilde{\Gamma}(i)
= \frac{2\epsilon}{\pi}
\left[
\tilde{\gamma} \frac{\hat{l}_2^2(i)}{\epsilon^2 + \hat{l}_1^2(i)}
+ \frac{\hat{l}_1^2(i)}{\epsilon^2 + \hat{l}_2^2(i)}
\right], 
\end{equation}
and $\tilde{\xi}(i)$ is calculated by Eq.~\eqref{tildexi}. 
\end{itemize}
For isotropic mobility,  simply  set $\tilde{M}=1$.

\subsection{Disconnections, stress, chemical potential jump, anisotropic energy and mobility}

We employ the following algorithm to evolve a chain of nodes in accordance with kinetic equation Eqs~(32) and (33) and employ 
the same reduced quantities as in Sect.~\ref{Purestepsanisotropicenergy},  
$\tilde{\tau} = \tau \alpha/\gamma^{(1)}$, and 
$\tilde{\psi} = \psi \alpha/\gamma^{(1)}$.  
\begin{itemize}
\item[(i)]
Initialise the coordinates of $N$ nodes: $\{\tilde{\mathbf{x}}(i)\}$ ($i = 0, \cdots, N-1$). 

\item[(ii)]
Displace each node by 
\begin{align}
&\left(
\def\arraystretch{1}
\begin{array}{c}
\tilde{x}_1(i)  \\
\tilde{x}_2(i) 
\end{array}\right)
:= 
\left(
\def\arraystretch{1}
\begin{array}{c}
\tilde{x}_1(i)  \\
\tilde{x}_2(i) 
\end{array}\right)
+ \Delta\tilde{t} 
\nonumber\\
&\times \left\{
\left[
\tilde{\Gamma}(i) \tilde{\kappa}(i) 
+ \tilde{\psi}
+ \tilde{\tau}(i) \Lambda
\right]
\left(
\def\arraystretch{1}
\begin{array}{c}
-\hat{l}_2(i)  \\
\tilde{M} \hat{l}_1(i) 
\end{array}\right)
+ 
\tilde{\xi}(i)
\left(
\def\arraystretch{1}
\begin{array}{c}
\hat{l}_1(i)  \\
\hat{l}_2(i) 
\end{array}\right)
\right\}, 
\end{align}
where $\tilde{\Gamma}(i)$ is obtained from Eq.~\eqref{tildeGammai} and $\tilde{\xi}(i)$ from Eq.~\eqref{tildexi}. 
The total shear stress is  
\begin{equation}
\tilde{\tau}(i)
= \tilde{G} \left[\beta^{(1)} \tilde{I}^{(1)}(i) + \beta^{(2)} \tilde{I}^{(2)}(i)\right]
+ \tilde{\tau}_\text{ext}, 
\end{equation}
where $\tilde{\tau}_\text{ext}$ is the (constant) externally applied shear stress, 
\begin{equation}
\tilde{I}^{(m)}(i)
= 
\sum_{j=0}^{N-1}
\tilde{x}_n'(j) \tilde{\tau}^{(m)}(i,j).
\end{equation}
Here, 
\begin{align}
&\tilde{\tau}^{(m)}(i,j)
= \dfrac{\tilde{x}_m(i) - \tilde{x}_m(j)}{\tilde{\varrho}_a^2(i,j)}
\left\{
1 - \dfrac{2[\tilde{x}_n(i) - \tilde{x}_n(j)]^2}{\tilde{\varrho}_a^2(i,j)}
\right\},  
\nonumber\\
&\tilde{\varrho}_a^2(i,j)
= [\tilde{x}_1(i) - \tilde{x}_1(j)]^2 + [\tilde{x}_2(i) - \tilde{x}_2(j)]^2 + \tilde{a}^2, 
\end{align}
and $\tilde{G} = G \alpha /[2\pi(1-\nu) \gamma^{(1)}]$. 
\end{itemize}
The  matrices $\tilde{\tau}^{(1)}(i,j)$ and $\tilde{\tau}^{(2)}(i,j)$m are $N\times N$ matrices (where $N$ may be large) have to be updated at each time step. 
We may reduce the computational by using  the fact that 
(i) we need to calculate is $\tilde{x}_k(i) - \tilde{x}_k(j)$ ($k=1,2$) and its square; and
(ii) $\tilde{\tau}^{(k)}(i,j) = - \tilde{\tau}^{(k)}(j,i)$ and $\tilde{\tau}^{(k)}(i,i) = 0$.

\section{Non-orthogonal reference interfaces}

\subsection{Two example cases}

\begin{figure*}[t]
\includegraphics[width=0.9\linewidth]{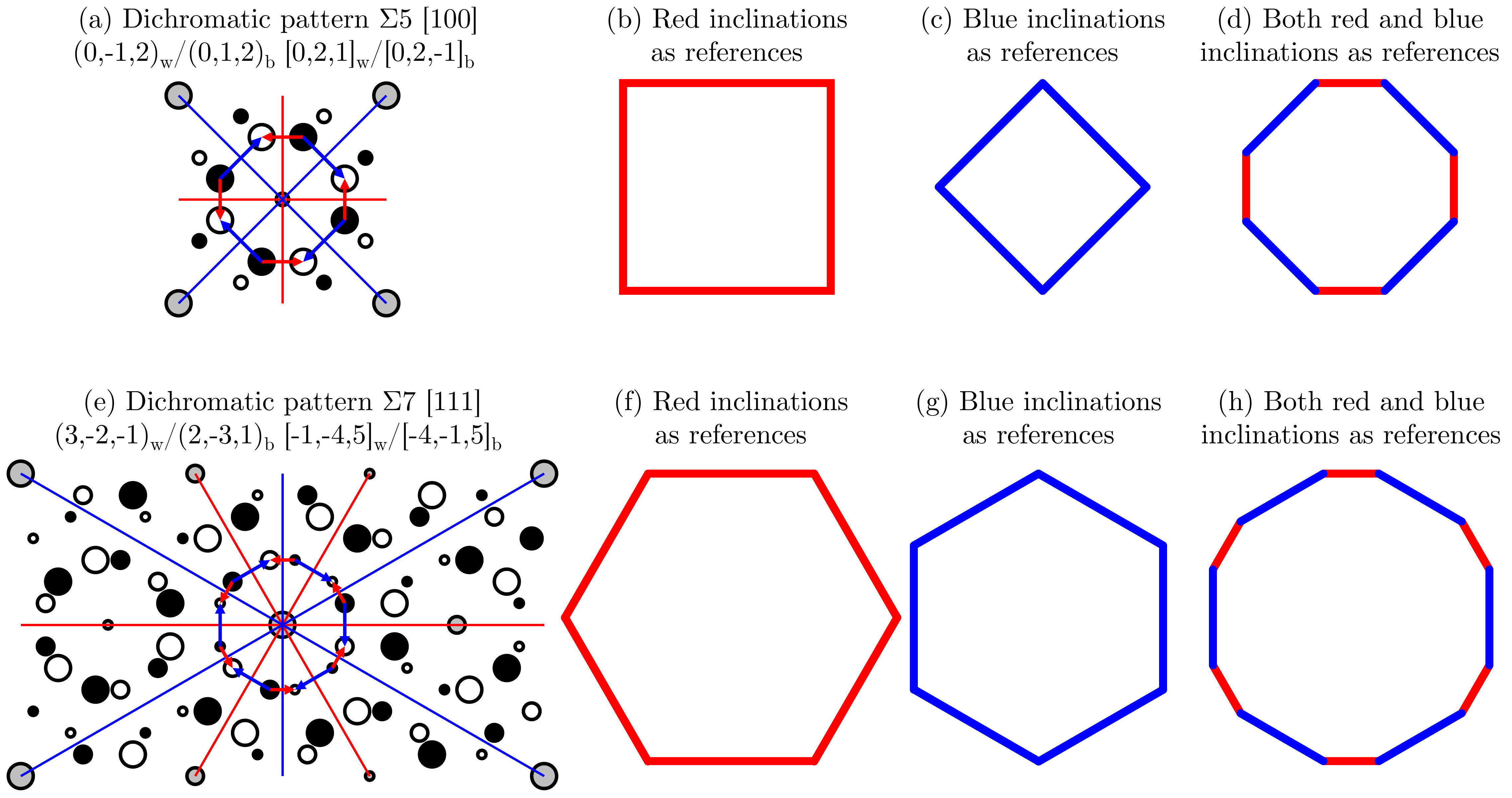}\hspace{-1.78em}%
\caption{\label{DSCs}
(a) Dichromatic pattern for a $\Sigma 5$ $[100]$ tilt GB in cubic crystals. 
Red and blue lines are  possible reference interface planes. 
Red(blue) arrows are  short DSC vectors parallel to the red(blue) reference interfaces. 
(b), (c) and (d) show the possible faceted GB profiles when we take the red inclinations, the blue inclinations, and both red and blue inclinations in (a) as  references. 
(e) As in (a) but for a $\Sigma 7$ $[111]$ tilt GB in cubic crystals. 
(f), (g) and (h) show  possible faceted GB profiles when we take the red inclinations, the blue inclinations, and both red and blue inclinations in (e) as  references. 
}
\end{figure*}

In Fig.~\ref{DSCs}, we show two examples for situation where the reference interfaces are not orthogonal. 
For a $\Sigma 5$ $[100]$ tilt GB (Fig.~\ref{DSCs}a), eight reference interfaces are available; see Fig.~\ref{DSCs}d. 
There are short DSC vectors parallel to each of these reference interfaces; see the arrows in Fig.~\ref{DSCs}d. 
For a $\Sigma 7$ $[111]$ tilt GB (Fig.~\ref{DSCs}e), 12 reference interfaces are available; see Fig.~\ref{DSCs}h. 
There are short DSC vectors parallel to each of these reference interfaces; see the arrows in Fig.~\ref{DSCs}e. 

\subsection{Geometric description}

Here, we illustrate the case of non-orthogonal reference interfaces. 
The two references for any interface segment are selected based on the direction of local tangent vector, as described in the text. 
Suppose that the potential reference interfaces are along the $\{\mathbf{e}^{(k)}\}$-axis. 
The inclination angle of each reference interface is $\phi^{(k)} = \arcsin(\mathbf{e}^{(k)} \cdot \mathbf{e}_1)$ (recall that $\mathbf{e}_1$ and $\mathbf{e}_2$ are the Cartesian coordinate system axes). 
For local tangent vector  $\hat{\mathbf{l}}$,  the local inclination angle is $\phi(s) = \arcsin(\hat{\mathbf{l}}(s) \cdot \mathbf{e}_1)$. 
If $\phi$ is in the range $[\phi^{(k)}, \phi^{(k+1)}]$, we  select   reference interfaces R($k$) and R($k+1$). 

For simplicity we assume that $\phi \in [\phi^{(1)}, \phi^{(2)}]$ and select R(1) and R(2) as the references (more general results are recovered with $(1,2)\rightarrow(k,k+1)$). 
Any vector may be represented in $(\mathbf{e}_1, \mathbf{e}_2)$ or $(\mathbf{e}^{(1)}, \mathbf{e}^{(2)})$ as 
\begin{align}
\mathbf{x}
&= x_1\mathbf{e}_1 + x_2\mathbf{e}_2
= x^{(1)}\mathbf{e}^{(1)} + x^{(2)}\mathbf{e}^{(2)}
\nonumber\\
&=
\left(\begin{array}{cc}
\mathbf{e}_1 & \mathbf{e}_2
\end{array}\right)
\left(\begin{array}{c}
x_1 \\ x_2
\end{array}\right)
=
\left(\begin{array}{cc}
\mathbf{e}^{(1)} & \mathbf{e}^{(2)}
\end{array}\right)
\left(\begin{array}{c}
x^{(1)} \\ x^{(2)}
\end{array}\right).
\end{align}
The transformation is 
\begin{equation}\label{transform}
\left(\begin{array}{c}
x_1 \\ x_2
\end{array}\right)
=
\left(\begin{array}{cc}
\mathbf{e}_1\cdot\mathbf{e}^{(1)} & \mathbf{e}_1\cdot\mathbf{e}^{(2)} \\
\mathbf{e}_2\cdot\mathbf{e}^{(1)} & \mathbf{e}_2\cdot\mathbf{e}^{(2)} 
\end{array}\right)
\left(\begin{array}{c}
x^{(1)} \\ x^{(2)}
\end{array}\right).
\end{equation}
Define $\rho^{(k)}(s)\ud L$ ($k = 1,2$) as the number of R($k$) disconnections in the segment $\ud L$. 
Geometric analysis gives 
\begin{subequations}
\begin{gather}
h^{(1)} \rho^{(1)} \ud L = \ud x^{(2)}, 
\end{gather}
\begin{gather}
-h^{(2)} \rho^{(2)} \ud L = \ud x^{(1)}.
\end{gather}
\end{subequations}
Combining this with Eq.~\eqref{transform} gives 
\begin{subequations}\label{tangenteehrho}
\begin{gather}
\hat{l}_1 
= \left(\mathbf{e}_1\cdot\mathbf{e}^{(2)}h^{(1)} \rho^{(1)}
- \mathbf{e}_1\cdot\mathbf{e}^{(1)}h^{(2)} \rho^{(2)}\right),
\end{gather}
\begin{gather}
\hat{l}_2
= \left(\mathbf{e}_2\cdot\mathbf{e}^{(2)}h^{(1)} \rho^{(1)}
- \mathbf{e}_2\cdot\mathbf{e}^{(1)}h^{(2)} \rho^{(2)}\right), 
\end{gather}
\end{subequations}
from which we  obtain   $h^{(1)} \rho^{(1)}$ and $h^{(2)} \rho^{(2)}$. 

\subsection{Curvature-flow kinetics}

\begin{figure}[bt!]
\includegraphics[height=0.55\linewidth]{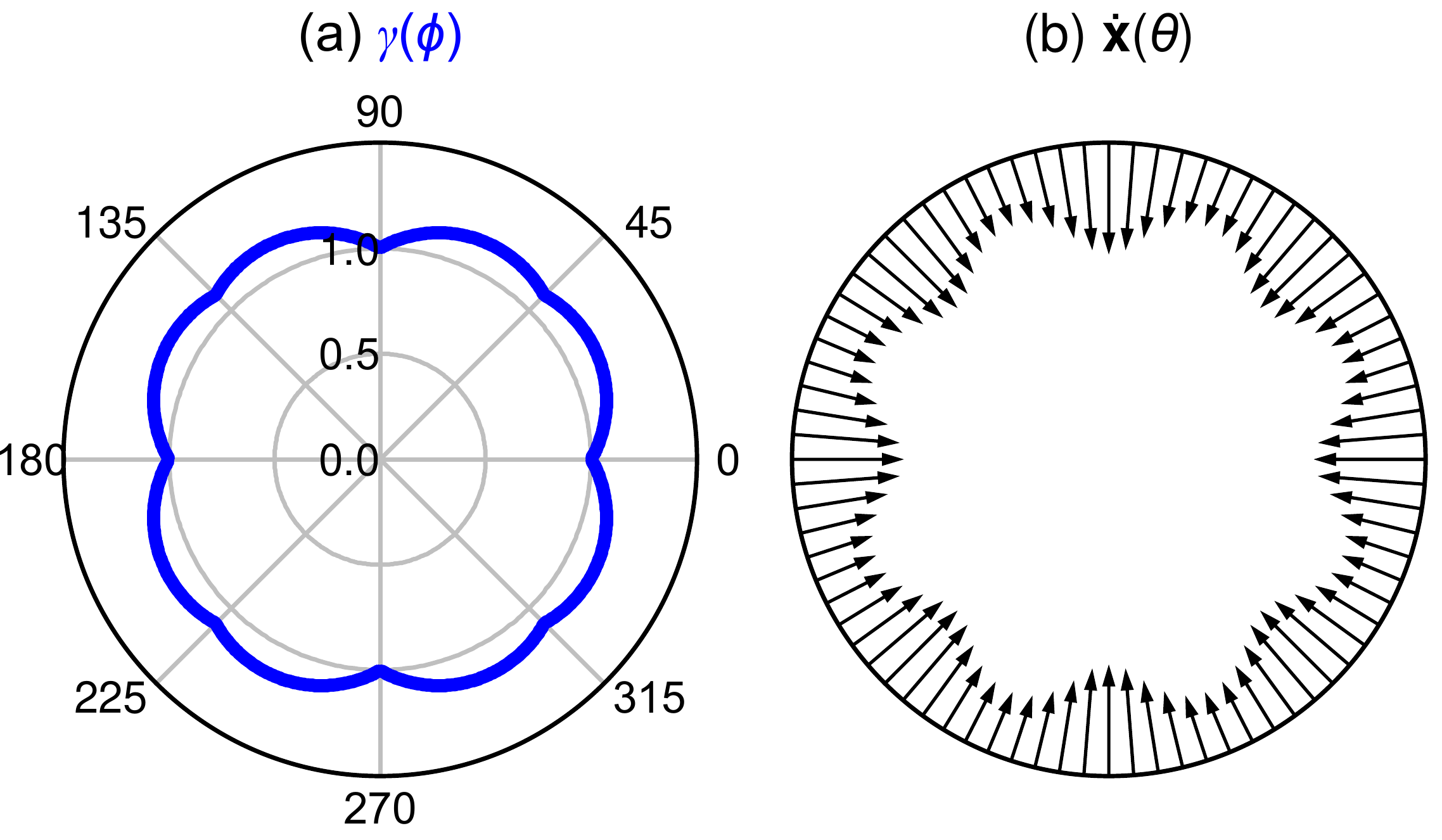}\hspace{-1.78em}%
\caption{\label{fig:anisotropy_gamma_nonorthogonal}(a) Polar plot of interface energy vs. inclination angle $\gamma(\phi)$ (scaled by $\gamma^{(2)}$) with $\gamma^{(1)}/\gamma^{(2)} = 1.1$. 
(b) Velocity field distributed along a circular interface, according to Eq.~\eqref{GammaNonorth} with $\epsilon = 0.3$ and $\gamma^{(1)}/\gamma^{(2)} = 1.1$. 
}
\end{figure}

In  the pure-step model, the anisotropic interface energy is 
\begin{equation}
\gamma 
= \gamma^{(2)}h^{(2)} |\rho^{(2)}|
+ \gamma^{(1)}h^{(1)} |\rho^{(1)}|. 
\end{equation}
Substituting $h^{(1)} \rho^{(1)}$ and $h^{(2)} \rho^{(2)}$ into this equation yields
\begin{align}
\gamma
&= 
\gamma^{(2)} \left|
\dfrac{\frac{\hat{l}_1}{\mathbf{e}_1\cdot\mathbf{e}^{(2)}} - \frac{\hat{l}_2}{\mathbf{e}_2\cdot\mathbf{e}^{(2)}}}
{\frac{\mathbf{e}_1\cdot\mathbf{e}^{(1)}}{\mathbf{e}_1\cdot\mathbf{e}^{(2)}} - \frac{\mathbf{e}_2\cdot\mathbf{e}^{(1)}}{\mathbf{e}_2\cdot\mathbf{e}^{(2)}}}
\right|
+
\gamma^{(1)} \left|
\dfrac{\frac{\hat{l}_1}{\mathbf{e}_1\cdot\mathbf{e}^{(1)}} - \frac{\hat{l}_2}{\mathbf{e}_2\cdot\mathbf{e}^{(1)}}}
{\frac{\mathbf{e}_1\cdot\mathbf{e}^{(2)}}{\mathbf{e}_1\cdot\mathbf{e}^{(1)}} - \frac{\mathbf{e}_2\cdot\mathbf{e}^{(2)}}{\mathbf{e}_2\cdot\mathbf{e}^{(1)}}}
\right|
\nonumber\\
&=
\gamma^{(2)} \left|
\dfrac{\sin(\phi^{(2)} - \phi)}{\sin(\phi^{(2)} - \phi^{(1)})}
\right|
+
\gamma^{(1)} \left|
\dfrac{\sin(\phi - \phi^{(1)})}{\sin(\phi^{(2)} - \phi^{(1)})}
\right|.
\end{align}
We confirm this expression for the special case of four reference interfaces, with  basis vectors $\{\mathbf{e}^{(1)}, \mathbf{e}^{(2)}, \mathbf{e}^{(3)}, \mathbf{e}^{(4)}\}$. 
Their inclination angles are  $n\pi/4$, $n\in\{0,1, \cdots, 7\}$. 
Figure~\ref{fig:anisotropy_gamma_nonorthogonal}a shows the interface energy vs. inclination angle for  $\gamma^{(1)} = \gamma^{(3)}$, $\gamma^{(2)} = \gamma^{(4)}$ and $\gamma^{(1)}/\gamma^{(2)} = 1.1$. 
Hence,  the interface stiffness ($\Gamma = \gamma + \gamma_{,\phi\phi}$) is
\begin{align}\label{GammaNonorth}
&\Gamma
= -M \frac{2\epsilon}{\pi |\sin(\phi^{(2)} - \phi^{(2)})|} 
\nonumber\\
&\times\left[
\gamma^{(2)}
\dfrac{\cos^2(\phi^{(2)} - \phi)}
{\epsilon^2 + \sin^2(\phi^{(2)} - \phi)}
+ 
\dfrac{\cos^2(\phi - \phi^{(1)})}
{\epsilon^2 + \sin^2(\phi - \phi^{(1)})}
\right], 
\end{align}
where we have applied the regularisation of Eq.~(11). 
The interface velocity is $\dot{\mathbf{x}} = M \Gamma \kappa \hat{\mathbf{n}}$, as shown in
Fig.~\ref{fig:anisotropy_gamma_nonorthogonal}b  for a circular interface of radius $R$. 

\subsection{Step-flow kinetics}

\begin{figure}[bt!]
\includegraphics[height=0.55\linewidth]{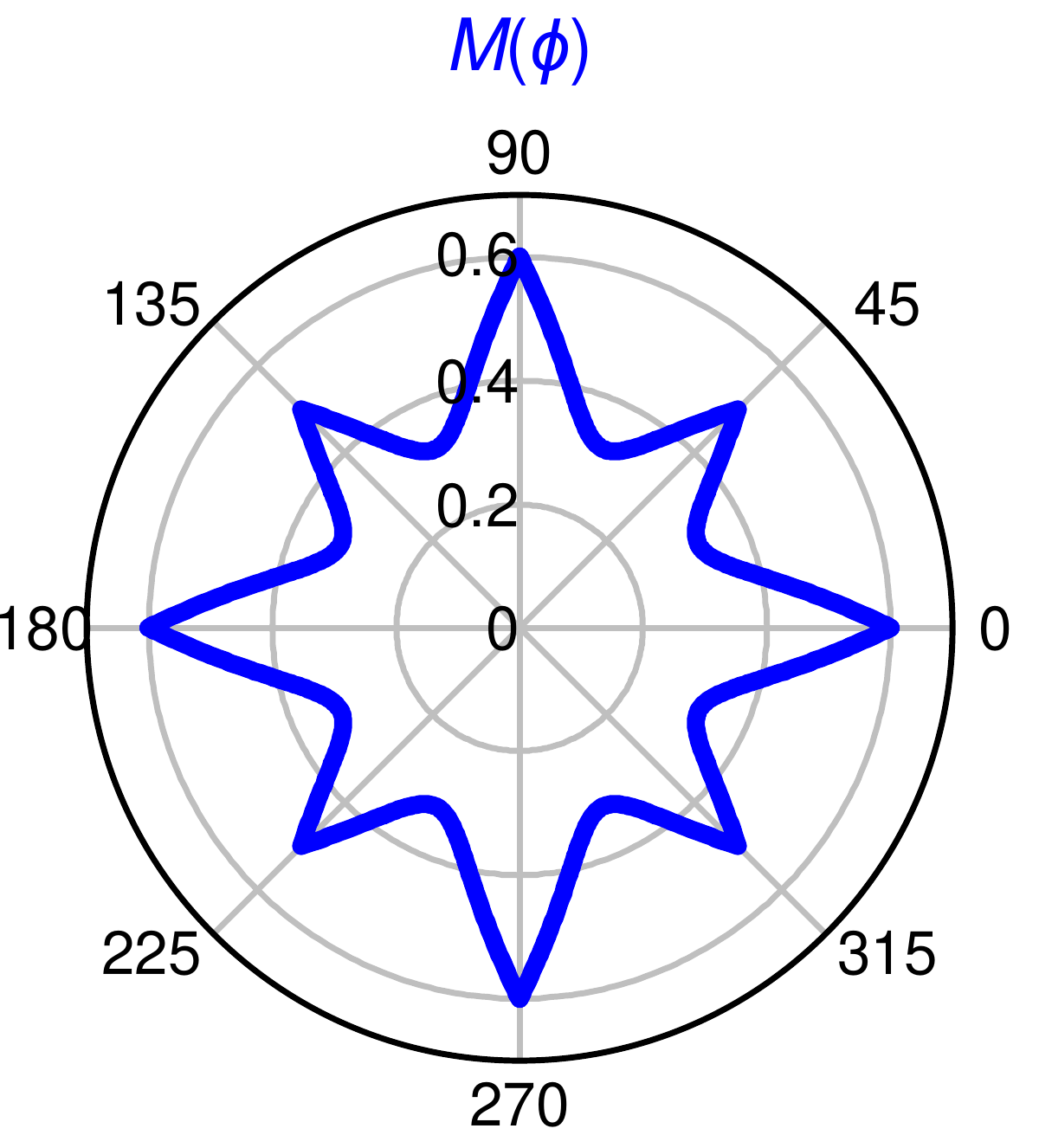}\hspace{-1.78em}%
\caption{\label{fig:anisotropy_kinetics_nonorthogonal}
Polar plot of the magnitude of interface mobility vs. inclination angle $M(\phi)$ (scaled by $M^{(1)}$) with $M^{(2)}/M^{(1)} = 1.2$. 
}
\end{figure}

In the dual-reference model, we  consider disconnection migrations on two interfaces (R(1) and R(2)) of inclination $\phi \in [\phi^{(1)}, \phi^{(2)}]$. 
The two directions  $\mathbf{e}^{(1)}$ and $\mathbf{e}^{(2)}$ need not beorthogonal. 
The variation of energy is 
\begin{align}
\delta E[\mathbf{x}]
&= \int \left(\Gamma\kappa\hat{l}_2 \delta x_1 
- \Gamma\kappa\hat{l}_1 \delta x_2\right) \ud L
\nonumber\\
&= \Gamma\kappa\int \bigg[
\left(\hat{l}_2\mathbf{e}_1\cdot\mathbf{e}^{(1)} - \hat{l}_1\mathbf{e}_2\cdot\mathbf{e}^{(1)}\right) \delta x^{(1)}
\nonumber\\
&+ \left(\hat{l}_2\mathbf{e}_1\cdot\mathbf{e}^{(2)} - \hat{l}_1\mathbf{e}_2\cdot\mathbf{e}^{(2)}\right) \delta x^{(2)}
\bigg] \ud L, 
\end{align}
where we have applied Eq.~\eqref{transform} to convert $\delta x_i$ to $\delta x^{(i)}$. 
The driving forces for the two disconnections types are
\begin{subequations}
\begin{gather}
f^{(1)}
= - \frac{\delta E}{\delta x^{(1)}}
= \Gamma\kappa
\left(\hat{l}_1\mathbf{e}_2\cdot\mathbf{e}^{(1)} - \hat{l}_2\mathbf{e}_1\cdot\mathbf{e}^{(1)}\right),
\end{gather}
\begin{gather}
f^{(2)}
= - \frac{\delta E}{\delta x^{(2)}}
= \Gamma\kappa
\left(\hat{l}_1\mathbf{e}_2\cdot\mathbf{e}^{(2)} - \hat{l}_2\mathbf{e}_1\cdot\mathbf{e}^{(2)}\right). 
\end{gather}
\end{subequations}
Assuming overdamped step migration, the interface velocity is 
\begin{align}\label{vGkMn}
\mathbf{v} 
&= v^{(1)} \mathbf{e}^{(1)} + v^{(2)} \mathbf{e}^{(2)}
= M^{(1)}f^{(1)} \mathbf{e}^{(1)} + M^{(2)}f^{(2)} \mathbf{e}^{(2)}
\nonumber\\
&= \Gamma \kappa \mathbf{M} \hat{\mathbf{n}}, 
\end{align}
where the mobility tensor is 
\begin{equation}\label{MMee}
\mathbf{M}
= M^{(1)} \mathbf{e}^{(1)}\otimes\mathbf{e}^{(1)}
+ M^{(2)} \mathbf{e}^{(2)}\otimes\mathbf{e}^{(2)}. 
\end{equation}
For orthogonal references (Eqs.~(16) and (17) in main text), the mobilities are simply Eqs.~\eqref{vGkMn} and \eqref{MMee} with $\phi^{(1)} = 0$ and $\phi^{(2)} = \pi/2$. 
The  flat interface mobility  is
\begin{equation}\label{Men}
M
= \hat{\mathbf{n}}\cdot\mathbf{M}\hat{\mathbf{n}}
= M^{(1)}(\mathbf{e}^{(1)}\cdot\hat{\mathbf{n}})^2
+ M^{(2)}(\mathbf{e}^{(2)}\cdot\hat{\mathbf{n}})^2. 
\end{equation}
By substituting $\hat{\mathbf{n}} = -\sin\phi\,\mathbf{e}_1 + \cos\phi\,\mathbf{e}_2$ into Eq.~\eqref{Men}, we  express $M$ as a function of the inclination angle:
\begin{equation}
M(\phi)
= M^{(1)}\sin^2(\phi - \phi^{(1)})
+ M^{(2)}\sin^2(\phi^{(2)} - \phi). 
\end{equation}
Figure~\ref{fig:anisotropy_kinetics_nonorthogonal} shows  $M(\phi)$ for $M^{(2)}/M^{(1)} = 1.2$. 
It is interesting to note that this plot is consistent with the molecular dynamics simulation results for a series of $\Sigma 5$ asymmetric tilt boundaries obtained by Lontine~\cite{SMlontine2018stress}. 
Because there are four reference boundaries for $\Sigma 5$ $[100]$ tilt boundaries in cubic crystals: $\mathbf{e}^{(1)} = [03\bar{1}]$, $\mathbf{e}^{(2)} = [021]$, $\mathbf{e}^{(3)} = [013]$, and $\mathbf{e}^{(4)} = [0\bar{1}2]$. 
We also note that when $M^{(2)}/M^{(1)}$ is large (e.g. $\sim 4$), the spikes in $M(\phi)$ at $\phi = n\pi/4$ are very small. 

\subsection{Disconnection model}

For interface R($k$), the disconnection Burgers vector   is $\mathbf{b}^{(k)} = b^{(k)}\mathbf{e}^{(k)}$. 
Under  stress $\boldsymbol{\upsigma}$, the Peach-Koehler (PK) force on  disconnection $k$ is 
\begin{equation}
\mathbf{f}^{(k)}_\text{PK}
= (\boldsymbol{\upsigma}\mathbf{b}^{(k)}) \times \mathbf{e}_3
= b^{(k)}\left(\begin{array}{c}
\sigma_{12}\cos\phi^{(k)} + \sigma_{22}\sin\phi^{(k)} \\
-\sigma_{11}\cos\phi^{(k)} - \sigma_{12}\sin\phi^{(k)} \\
0
\end{array}\right).
\end{equation}
The glide driving force  is 
\begin{equation}
f_\ug^{(k)}
= \mathbf{f}^{(k)}_\text{PK} \cdot \mathbf{e}^{(k)}
= \tau^{(k)} b^{(k)}, 
\end{equation}
where the stress  resolved onto the R($k$) interface is 
\begin{equation}
\tau^{(k)}
= \frac{\sigma_{22} - \sigma_{11}}{2} \sin 2\phi^{(k)}
+ \sigma_{12} \cos 2\phi^{(k)}. 
\end{equation}

For the  R(1) and R(2) references, 
Type~(1) migration the energy variation with R(1) disconnection displacement is 
\begin{equation}
\delta E^{(1)}
= -\tau^{(1)} \mathcal{B}^{(1)} h^{(1)}\rho^{(1)} \delta g^{(1)}, 
\end{equation}
where $\mathcal{B}^{(1)} \equiv b^{(1)}/h^{(1)}$. 
Note, $\mathcal{B}^{(1)}$ is not the shear-coupling factor~\cite{SMcahn2006coupling} (i.e., the shear-coupling factor is the ratio between the Burgers vector and the orthogonal component of the step height); rather it is related to  the shear-coupling factor by $\beta^{(1)} = \mathcal{B}^{(1)}/\sin(\phi^{(2)} - \phi^{(1)})$.  
For the coupled motion of the R(2) disconnection, we have 
\begin{equation}
\delta E_\text{coupling}^{(2)}
= \tau^{(2)} \mathcal{B}^{(2)} h^{(1)}\rho^{(1)} \delta g^{(1)}. 
\end{equation}
The change in the total energy is $\delta E = \delta E^{(1)} + \delta E_\text{coupling}^{(2)}$ and the driving force on Type~(1) migration is 
\begin{equation}\label{fg1}
f_\ug^{(1)}
= - \frac{\delta E}{\delta g^{(1)}}
= -\left(\tau^{(2)} \mathcal{B}^{(2)} - \tau^{(1)} \mathcal{B}^{(1)}\right)
 h^{(1)}\rho^{(1)}. 
\end{equation}
Similarly, the driving force on Type~(2) migration is 
\begin{equation}\label{fg2}
f_\ug^{(2)}
= -\left(\tau^{(2)} \mathcal{B}^{(2)} - \tau^{(1)} \mathcal{B}^{(1)}\right)
 h^{(2)}\rho^{(2)}. 
\end{equation}

If we only consider the stress as the driving force (ignore  capillarity), the interface velocity is 
\begin{equation}\label{vMfge}
\mathbf{v}
= M^{(1)} f_\ug^{(1)} \mathbf{e}^{(1)}
+ M^{(2)} f_\ug^{(2)} \mathbf{e}^{(2)}. 
\end{equation}
Substituting Eq.~\eqref{fg1}, Eq.~\eqref{fg2}, $h^{(1)}\rho^{(1)}$ and $h^{(2)}\rho^{(2)}$ (from Eq.~\eqref{tangenteehrho}) into Eq.~\eqref{vMfge}, we finally obtain:
\begin{equation}
\mathbf{v}
= \boldsymbol{\tau} \cdot \boldsymbol{\lambda} \mathbf{M} \hat{\mathbf{n}}, 
\end{equation}
where $\boldsymbol{\tau}$ is the vector of resolved shear stresses
\begin{equation}
\boldsymbol{\tau}
= \left(\begin{array}{c}
\tau^{(1)} \\ \tau^{(2)}
\end{array}\right),
\end{equation}
$\boldsymbol{\lambda}$ is the shear-coupling factor vector
\begin{equation}
\boldsymbol{\lambda}
= \left(\begin{array}{c}
-\beta^{(1)} \\ \beta^{(2)}
\end{array}\right)
= \frac{1}{\sin(\phi^{(2)} - \phi^{(1)})}
\left(\begin{array}{c}
-\mathcal{B}^{(1)} \\ \mathcal{B}^{(2)}
\end{array}\right),
\end{equation}
and $\mathbf{M}$ is the mobility tensor defined in Eq.~\eqref{MMee}. 
Including capillarity contributions, the kinetic equation is 
\begin{equation}
\mathbf{v}
= \left(\Gamma\kappa + \boldsymbol{\tau} \cdot \boldsymbol{\lambda}\right) 
\mathbf{M} \hat{\mathbf{n}}. 
\end{equation}

If a chemical potential jump $\psi$ is also present, it is easy to show that the interface velocity driven by $\psi$ is $\mathbf{v} = \psi \mathbf{M}\hat{\mathbf{n}}$. 
Hence, the general kinetic equation for the motion of an interface is 
\begin{equation}
\mathbf{v} = \mathbf{M} \mathbf{f}, 
\end{equation}
where $\mathbf{M}$ is as per Eq.~\eqref{MMee} and $\mathbf{f}$ is the total driving force
\begin{equation}
\mathbf{f}
\equiv \left(\Gamma\kappa + \psi + \boldsymbol{\tau} \cdot \boldsymbol{\lambda}\right)\hat{\mathbf{n}}. 
\end{equation}

\section{Capillarity vector}

Here, we will show the equivalence between the weighted mean curvature $\Gamma \kappa$ and the interface divergence of the capillarity vector $\nabla\cdot\boldsymbol{\xi}$ along a 1D interface curve. 
The capillarity vector is defined as [D.W. Hoffman and J.W. Cahn,  Surface Science {\bf 31}, 368–388 (1972)]
\begin{equation}
\boldsymbol{\xi}
= \nabla(r\gamma)
= \gamma \hat{\mathbf{n}} + \gamma_{,\phi}\mathbf{e}_\phi
= \gamma \hat{\mathbf{n}} - \gamma_{,\phi}\hat{\mathbf{l}}, 
\end{equation}
and its divergence is  
\begin{equation}\label{divxi}
\nabla\cdot\boldsymbol{\xi}
= \frac{\partial\boldsymbol{\xi}}{\partial L}\cdot \mathbf{e}_\phi
= -\frac{1}{|\mathbf{l}|} \boldsymbol{\xi}'\cdot \hat{\mathbf{l}},
\end{equation}
where $\hat{\mathbf{l}} = -\mathbf{e}_\phi$ and
\begin{equation}\label{xipl}
\boldsymbol{\xi}' \cdot \hat{\mathbf{l}}
= \left\{\frac{1}{|\mathbf{l}|}\left[
\gamma\left(\begin{array}{c}
-x_2' \\ x_1'
\end{array}\right)
- \gamma_{,\phi}\left(\begin{array}{c}
x_1' \\ x_2'
\end{array}\right)
\right]\right\}'
\cdot
\frac{1}{|\mathbf{l}|}
\left(\begin{array}{c}
x_1' \\ x_2'
\end{array}\right).
\end{equation}

The first term in this  expression is 
\begin{align}
&\left(\frac{1}{|\mathbf{l}|}\right)'
\left[
\gamma\left(\begin{array}{c}
-x_2' \\ x_1'
\end{array}\right)
- \gamma_{,\phi}\left(\begin{array}{c}
x_1' \\ x_2'
\end{array}\right)
\right]
\cdot
\frac{1}{|\mathbf{l}|}
\left(\begin{array}{c}
x_1' \\ x_2'
\end{array}\right)
\nonumber\\
&=
\gamma_{,\phi}\frac{|\mathbf{l}|'}{|\mathbf{l}|}
=
\gamma_{,\phi} \frac{x_1'x_1'' + x_2' x_2''}{x_1'^2 + x_2'^2}. 
\end{align}
The second term in Eq.~\eqref{xipl} is 
\begin{align}
&\frac{1}{|\mathbf{l}|}
\left[
\gamma\left(\begin{array}{c}
-x_2' \\ x_1'
\end{array}\right)
- \gamma_{,\phi}\left(\begin{array}{c}
x_1' \\ x_2'
\end{array}\right)
\right]'
\cdot
\frac{1}{|\mathbf{l}|}
\left(\begin{array}{c}
x_1' \\ x_2'
\end{array}\right)
\nonumber\\
&=
\gamma \frac{x_2' x_1'' - x_1' x_2''}{x_1'^2 + x_2'^2}
- \gamma_{,\phi}'
- \gamma_{,\phi} \frac{x_1'x_1'' + x_2' x_2''}{x_1'^2 + x_2'^2}
\nonumber\\
&=
-\gamma \kappa |\mathbf{l}|
- \gamma_{,\phi}'
- \gamma_{,\phi} \frac{x_1'x_1'' + x_2' x_2''}{x_1'^2 + x_2'^2}.
\end{align}
So, Eq.~\eqref{xipl} becomes 
\begin{equation}\label{xipl2}
\boldsymbol{\xi}'\cdot\hat{\mathbf{l}}
= -\left(\gamma\kappa|\mathbf{l}| + \gamma_{,\phi}'\right), 
\end{equation}
where 
\begin{equation}\label{gammapphi}
\gamma_{,\phi}'
= \gamma_{,\phi\phi}\phi'
= \gamma_{,\phi\phi}\left(
\frac{\partial\phi}{\partial x_1'}x_1''
+ \frac{\partial\phi}{\partial x_2'}x_2''
\right)
= \gamma_{,\phi\phi} \kappa |\mathbf{l}|, 
\end{equation}
where we have used the relationship $\phi = \arctan(x_2'/x_1')$. 
Substituting Eq.~\eqref{gammapphi} into Eq.~\eqref{xipl2} and considering Eq.~\eqref{divxi}, we have 
\begin{equation}
\nabla\cdot\boldsymbol{\xi} = (\gamma + \gamma_{,\phi\phi})\kappa
= \Gamma\kappa.
\end{equation}

\end{document}